\documentclass[aip,amsmath,amssymb,reprint]{revtex4-1}
\usepackage{graphicx}
\usepackage{dcolumn}
\usepackage{bm}

\usepackage[utf8]{inputenc}
\usepackage[T1]{fontenc}
\usepackage{mathptmx}
\usepackage{etoolbox}

\usepackage{color}

\makeatletter
\def\@email#1#2{%
 \endgroup
 \patchcmd{\titleblock@produce}
  {\frontmatter@RRAPformat}
  {\frontmatter@RRAPformat{\produce@RRAP{*#1\href{mailto:#2}{#2}}}\frontmatter@RRAPformat}
  {}{}
}%

\newcommand*{\citen}[1]{%
	\begingroup
	\romannumeral-`\x 
	\setcitestyle{numbers}%
	\cite{#1}%
	\endgroup
}

\makeatother

\begin{document}
\title{Macroscopic behavior of populations of quadratic integrate-and-fire neurons subject to non-Gaussian white noise}
\author{Denis S.\ Goldobin}
\affiliation{Institute of Continuous Media Mechanics, UB RAS, Academician Korolev
Street 1, 614013 Perm, Russia}
\affiliation{Department of Control Theory, Nizhny Novgorod State University, Gagarin Avenue 23, 603022 Nizhny Novgorod, Russia}
\author{Evelina V.\ Permyakova}
\affiliation{Institute of Continuous Media Mechanics, UB RAS, Academician Korolev
Street 1, 614013 Perm, Russia}
\author{Lyudmila S.\ Klimenko}
\affiliation{Institute of Continuous Media Mechanics, UB RAS, Academician Korolev
Street 1, 614013 Perm, Russia}
\affiliation{Department of Theoretical Physics, Perm State University, Bukirev
Street 15, 614990 Perm, Russia}
\email{lyudmilaklimenko@gmail.com}
\date{\today}

\begin{abstract}
We study macroscopic behavior of populations of quadratic integrate-and-fire neurons subject to non-Gaussian noises; we argue that these noises must be alpha-stable whenever they are delta-correlated (white). For the case of additive-in-voltage noise, we derive the governing equation of the dynamics of the characteristic function of the membrane voltage distribution and construct a linear-in-noise perturbation theory. Specifically for the recurrent network with global synaptic coupling, we theoretically calculate the observables: population-mean membrane voltage and firing rate. The theoretical results are underpinned by the results of numerical simulation for homogeneous and heterogeneous populations.
The possibility of the generalization of the pseudocumulant approach to the case of a fractional $\alpha$ is examined for both irrational and fractional rational $\alpha$. This examination seemingly suggests the pseudocumulant approach or its modifications to be employable only for the integer values of $\alpha=1$ (Cauchy noise) and $2$ (Gaussian noise) within the physically meaningful range $(0;2]$. Remarkably, the analysis for fractional $\alpha$ indirectly revealed that, for the Gaussian noise, the minimal asymptotically rigorous model reduction must involve three pseudocumulants and the two-pseudocumulant model reduction is an artificial approximation. This explains a surprising gain of accuracy for the three-pseudocumulant models as compared to the the two-pseudocumulant ones reported in the literature.
\end{abstract}

\maketitle

\begin{quotation}
The abundance of a surprisingly low effective dimensionality of macroscopic dynamics of classical model oscillator populations had been puzzling researchers for a long time before it received a proper mathematical explanation. The Watanabe--Strogatz and Ott--Antonsen theories elucidated the laws of these dynamics for an important class of systems. With individual (intrinsic) noise the attractivity of low dimensional regimes is often even more pronounced, but the  Ott--Antonsen theory is inapplicable. The mathematical key to the solution of this problem
came with the circular cumulant formalism. For the model of quadratic integrate-and-fire neuron, which is the normal form for the Class~I excitability neurons near the bifurcation point, Montbri\'o, Paz\'o, and Roxin offered an alternative framework. It is mathematically equivalent to the Ott--Antonsen theory, but employs generic observables: population-mean membrane potential and firing rate. The formalism of pseudocumulants provided a language for dealing with violations of the  Montbri\'o--Paz\'o--Roxin theory and, specifically, the case of Gaussian noise.
However, the effective noise generated by fluctuations is not necessarily Gaussian and can be alpha-stable, i.e., its distribution can possess heavy power-law tails.
We look at the possibility to generalize the Montbri\'o--Paz\'o--Roxin framework to the case of fractional alpha-stable noises via the pseudocumulant formalism and the dynamics of the characteristic function.
While our mathematical results can be directly transferred to the problem of the Anderson localization in 1D setups and some other condensed matter problems, for the sake of specificity, we focus on the interpretations linked to the population of quadratic integrate-and-fire neurons with global synaptic coupling.
\end{quotation}

\section{Introduction}
In collective dynamics of populations of neurons, fluctuations play an essential role.~\cite{Geisler-Brunel-Wang-2005} These fluctuations can be both extrinsic and endogenous; they can be often  represented as an effective noise. Noteworthily, many noise-induced collective phenomena, like coherent oscillations, can be reproduced~\cite{Volo-Torcini-2018,Ratas-Pyragas-2019,Goldobin-Volo-Torcini-2021,Volo-etal-2022,Zheng-Kotani-Jimbo-2021} with the basic mathematical model of a quadratic integrate-and-fire (QIF) neuron.~\cite{Ermentrout-Kopell-1986,Izhikevich-2007}

Strictly saying, by virtue of the Central limit theorem, the superposition of a large number of sources of {\em finite} fluctuations can generate only Gaussian distributions. However, the separation of time scales of different Gaussian components can result in the necessity to introduce several effective Gaussian noises with non-stationary properties or some other complications. Meanwhile, in diverse limiting cases, mathematical idealizations result in a single non-Gaussian ($\alpha$-stable) noise. For many real-life systems the idealization of $\alpha$-stable distribution also provides an accurate description with only a few quantifiers, while the representation by a superposition of Gaussian components creates an obscure picture with a plenty of parameters and large uncertainties in their values due to the imperfectness and finiteness of any real-life statistical data sets.

Thus, some real fluctuation sources can be either adequately represented by a complex superposition of Gaussian noises with sophisticated separation of time scales, or better described by an idealization of a single stationary noise with heavy power-law tails~\cite{Petrovskii-Morozov-2009,Petrovskii-etal-2011} $\propto 1/|x|^{\alpha+1}$, an $\alpha$-stable noise, for which we will provide a moderately detailed introduction in the section with mathematical preliminaries. In particular, the synchronization of limit cycle oscillators by common noise~\cite{Pikovsky-1984,Teramae-Tanaka-2004,Goldobin-Pikovsky-2005a} in imperfect situations results in a Cauchy distribution of oscillator deviations from the cluster center;~\cite{Goldobin-Pikovsky-2005b} this corresponds to $\alpha=1$. Later on,~\cite{Goldobin-Dolmatova-2019a} in the limit of high synchrony, it was found that the interplay of the mechanisms of synchronization by coupling and common noise generates $\alpha$-stable fluctuations with $\alpha<1$ for a repulsive coupling and $\alpha>1$ for an attractive one. Furthermore, the synchronization by common noise can be also viewed as a theoretical framework for the generalized synchronization of chaotic oscillations,~\cite{Rulkov-etal-1995,Rosenblum-etal-1996,Pikovsky-Rosenblum-Kurths-2001-2003} where the driving chaotic signal is considered as a colored noise.~\cite{Nakao-etal-2010} Summarizing, the importance of $\alpha$-stable fluctuations in large ensembles of coupled oscillators with {\em slightly imperfect} synchrony can hardly be underestimated. If the effective endogenous noise in the network of coupled QIFs emerges from the asynchronous dynamics of individual elements, one can expect the existence of setups (network structures) where this noise can be $\alpha$-stable with $\alpha\ne2$ (non-Gaussian).

In 2008, Ott and Antonsen (OA) constructed a theory which gives an exact self-contained dynamic equation for the Kuramoto order parameter for an important class of phase oscillators.~\cite{Ott-Antonsen-2008,Ott-Antonsen-2009} For QIFs, one can switch from a membrane potential $V$ to a phase-like variable $\theta$ via the transformation $\theta=2\arctan{V}$ and find a phase dynamics equation of the sort for which the OA theory is valid. On the basis of the OA theory a ``next generation of neuron mass models''~\cite{Coombes-2023} was constructed.~\cite{Luke-Barreto-So-2013,Pazo-Montbrio-2014,Laing-2014,Laing-2018} Specifically for QIFs, Montbri\'o, Paz\'o, and Roxin (MPR) discovered~\cite{Montbrio-Pazo-Roxin-2015} that if one deals with $V$ without switching to $\theta$, one can derive an even more simple self-contained dynamic equation system for two generic observables: the population-mean membrane potential and the instantaneous firing rate. Mathematically, the OA and MPR theories correspond to the dynamics on the same manifold (though in different variables), but these approaches offer different primary observables and, more fundamentally, require different perturbative approaches, where their applicability conditions are imperfectly met. Both these theories are inapplicable when the elements are subject to nonidentical noise signals. One can make their generalization for the case of a Cauchy noise,~\cite{Toenjes-Pikovsky-2020,Pietras-etal-2023,Pyragas2-2023,Clusella-Montbrio-2022} but not any other noise.

For Gaussian noise, an advance with the theoretical macroscopic description~\cite{Ratas-Pyragas-2019,Zheng-Kotani-Jimbo-2021,Volo-etal-2022,Goldobin-2021} was made possible by the development of the formalism of so-called ``circular cumulants''~\cite{Tyulkina-etal-2018,Goldobin-Dolmatova-2019b}
which was introduced for the generalization of the OA theory beyond its original applicability conditions. While the circular cumulant formalism is a natural framework for the OA theory, specifically for QIFs an alternative ``pseudocumulant'' formalism~\cite{Goldobin-Volo-Torcini-2021} offers a generalization of the MPR theory.~\cite{Montbrio-Pazo-Roxin-2015} Within the latter formalism, the firing rate and the population-mean membrane voltage remain the primary dynamical variables entering the governing equations.

Recently, the cases of non-Gaussian stable noises acting on QIFs started to attract the attention in mathematical neuroscience. However, currently, this interest is limited to the only exactly solvable case of a Cauchy noise.~\cite{Pietras-etal-2023,Pyragas2-2023} The circular cumulant formalism was found useful for dealing with non-Gaussian noises,~\cite{Dolmatova-Tyulkina-Goldobin-2023} but specifically for the case of QIFs the more recent formalism of pseudocumulants can be even more promising. In this paper, we explore the possibility of the implementation/generalization of the pseudocumulant approach for/to the populations of QIFs subject to $\delta$-correlated non-Gaussian noise.

The paper is organized as follows. In Sec.~\ref{sec:Math}, we provide mathematical preliminaries: a brief introduction for the $\alpha$-stable distributions and $\delta$-correlated non-Gaussian noises, the fractional Fokker--Planck equation for additive noise. Further, we formulate the fractional Fokker--Planck description for the macroscopic dynamics of the recurrent synaptic network of quadratic integrate-and-fire neurons subject to non-Gaussian noise in Sec.~\ref{ssec:popQIF}. In Sec.~\ref{ssec:CF&PC}, we derive the governing equation for the dynamics of the characteristic function of the membrane voltage distribution and present the pseudocumulant formalism.
In Sec.~\ref{sec:fracA}, for the case of noninteger $\alpha$, we construct a first-order perturbation theory for the effect of noise on the characteristic function and derive macroscopic observables: population-mean voltage and firing rate. In Sec.~\ref{sec:QIFs}, the theoretical results for macroscopic states of homogeneous populations of QIFs are reported.
In Sec.~\ref{sec:disc}, we compare the theoretical results against the background of the results of numerical simulation (also for a less mathematically challenging case of heterogeneous populations) and discuss general implications of the theoretical results and the examination (Appendix~\ref{app:HIA}) of the possibility to construct the generalization of the pseudocumulant expansion to the case of noninteger $\alpha$. In Sec.~\ref{sec:conc}, conclusions are drawn.

\section{Mathematical preliminaries and formulation of the problem}
\label{sec:Math}

\subsection{Delta-correlated non-Gaussian noises}
\label{ssec:DCnG}
Let us consider a stochastic dynamical system subject to an additive noise;
\begin{equation}
\dot{\mathbf{x}}=\mathbf{f}(\mathbf{x})+\mathbf{g}\,\xi(t)\;,
\label{eq001}
\end{equation}
where $\mathbf{x}=\{x_1,x_2,...,x_M\}$; the properties of $\delta$-correlated noise $\xi(t)$ will be specified below, generally it is non-Gaussian.

For numerical simulation and a consistent theoretical formulation of the problem, it is instructive to consider the discrete-time version of Eq.~(\ref{eq001}) and the limit of time step size $\Delta t\to0$. In discrete time, we interpret Eq.~(\ref{eq001}) as
\begin{equation}
\mathbf{x}(t+\Delta t)=\mathbf{x}(t)
 +\left[\mathbf{f}\big(\mathbf{x}(t)\big)
 +\mathbf{g}\,\xi_{\Delta t}(t)\right]\Delta t\;.
\label{eq002}
\end{equation}
For a discrete-time version $\xi_{\Delta t,j}=\xi_{\Delta t}(t_j)$ of noise $\xi(t)$, the autocorrelation $\langle\xi_{\Delta t,j}\xi_{\Delta t,l\ne j}\rangle=0$, where $\langle\dots\rangle$ indicates averaging over the noise realizations. The normalization of $\xi_{\Delta t,j}$ for a non-Gaussian noise requires a subtle treatment and will be defined below.

In real-world systems, the probability density of fluctuations often possesses heavy tails;~\cite{Zolotarev-1986} in particular, this property is fundamentally inherent to fluctuations in oscillator ensembles synchronized by common noise.~\cite{Goldobin-Pikovsky-2005b,Goldobin-Dolmatova-2019a}
Let us consider Eq.~(\ref{eq002}) with time step $\Delta t$ and probability density of $\xi_{\Delta t}$ with tails $\propto1/|\xi_{\Delta t}|^{1+\alpha}$.
If we consider the system dynamics on the time scale $\tau_N=N\Delta t$, which is large compared to $\Delta t$ ($N\gg1$) but still small compared to the reference time scale of the noise-average dynamics ($\tau_N\ll1$), the noise will act as $\xi_{N\Delta t}=N^{-1}\sum_{j=1}^{N}\xi_{\Delta t,j}$ with independent $\xi_{\Delta t,j}$.
By virtue of the generalized central limit theorem, such sums are distributed according to a Gaussian distribution for $\alpha\ge 2$ and according to the $\alpha$-stable distributions for $0<\alpha<2$ (in the latter case, the distribution of $\xi_{N\Delta t}$ possesses $1/|\xi_{N\Delta t}|^{1+\alpha}$-tails). Thus, in numerical simulations, picking-up an $\alpha$-stable distribution will yield better convergence of the results to the limit $\Delta t\to0$. In theoretical analysis, a physically consistent consideration must employ the $\alpha$-stable statistics for a delta-correlated noise.

Now we briefly recall the properties of $\alpha$-stable distributions.~\cite{Zolotarev-1986}
For $\alpha\le0$ the distribution cannot be normalized. Generally, these distributions are defined via their characteristic functions;
for a random number $\xi$,
\begin{equation}
\langle{e^{ik\xi}}\rangle\equiv
F_\xi(k;\alpha,\beta,c,\mu)
=e^{ik\mu-|ck|^\alpha(1+i\beta\mathrm{sign}(k)\Theta)},
\label{eq003}
\end{equation}
where $\mu\in\mathbb{R}$ is a shift parameter, $c>0$ is a scale parameter featuring the distribution width, $\beta\in[-1,1]$ called the skewness parameter (typically the term ``skewness'' is related to the third cumulant, which diverges here, for $\alpha<2$),
\[
\Theta=\left\{
\begin{array}{cc}
\tan\left(\frac{\pi\alpha}{2}\right)\;, & \mbox{ for } \alpha\ne1\;; \\
-\frac{2}{\pi}\ln|k|\;, & \mbox{ for } \alpha=1\;.
\end{array}
\right.
\]
In particular, for a Cauchy random variable, $\alpha=1$, $\mu$ is the distribution median, $\beta=0$, and $c$ is a half-width at half-maximum; for a Gaussian random variable, $\alpha=2$, $\mu$ is the mean value, $\beta$ is multiplied by $0$, and the variance is $2c^2$.

It is important to discuss the scaling properties of the discrete-time version of a $\delta$-correlated $\alpha$-stable noise with $\Delta t$. The sum of two independent $\alpha$-stable random variables is an $\alpha$-stable random variable with parameters [see Eq.~(\ref{eq003})] $\mu=\mu_1+\mu_2$,
\begin{equation}
|c|=(|c_1|^\alpha+|c_2|^\alpha)^{1/\alpha},
\label{eq003c}
\end{equation}
$\beta=(\beta_1|c_1|^\alpha+\beta_2|c_2|^\alpha)/(|c_1|^\alpha+|c_2|^\alpha)$. For the sum of two noise increments $\xi_{\Delta{t}}\Delta{t}$ the equivalent increment for the time step size $(2\Delta{t})$ is $\xi_{2\Delta{t}}2\Delta{t}=2^{1/\alpha}\xi_{\Delta{t}}\Delta{t}$ [see the rule for the summation of amplitudes (\ref{eq003c})]; therefore,
\[
\xi_{\Delta{t}}\propto(\Delta{t})^{1/\alpha-1}.
\]
In particular, for a $\delta$-correlated Cauchy noise ($\alpha=1$) of amplitude $\sigma$, one should take characteristic function~(\ref{eq003}) of $\xi_{\Delta t}$ with $c=\sigma$. For a Gaussian noise ($\alpha=2$) of intensity $\sigma^2$, which corresponds to the continuous-time noise with $\langle\xi(t)\,\xi(t+t')\rangle=2\sigma^2\delta(t')$, one should take characteristic function~(\ref{eq003}) of $\xi_{\Delta t}$ with $c^2=\sigma^2/\Delta t$.

\subsection{Fractional Fokker-Planck equation for alpha-stable noise}
\label{ssec:fracFP}
For the stochastic system (\ref{eq001}) with additive noise, the probability density function $w(\mathbf{x},t)$ obeys a fractional Fokker-Planck equation:~\cite{Klyatskin-1980,Klyatskin-2005}
\begin{equation}
\frac{\partial w}{\partial t}
 +\sum_{l=1}^{M}\frac{\partial}{\partial x_l}\big(f_l(\mathbf{x})w\big)
 -\dot\Phi_t^{(\xi)}(i\hat{Q})w=0\,,
\label{eq105}
\end{equation}
where $\hat{Q}\equiv\sum_{l=1}^{M}g_l(\partial/\partial x_l)$.
For the case of symmetric $\alpha$-stable noise [Eq.~(\ref{eq003}) with $\mu=\beta=0$] of amplitude $\sigma$, function $\dot\Phi_t^{(\xi)}(k)=-\sigma^\alpha|k|^\alpha$ [see Appendix~\ref{app:FFP} for a minimalistic derivation and Eq.~(\ref{eq006}) for the intuition on notations]. A more general derivation can be found in Refs.~\citen{Klyatskin-1980,Klyatskin-2005} or Secs.~III.A and B of Ref.~\citen{Dolmatova-Tyulkina-Goldobin-2023}; e.g., Refs.~\citen{Chechkin-etal-2003,Toenjes-etal-2013} also deal with Eq.~(\ref{eq105}).
Here and hereafter, we consider only this symmetric case and use the form of operator $\dot\Phi_t^{(\xi)}(i\hat{Q})$ in the Fourier space, where $\dot\Phi_t^{(\xi)}\left(i\sum_{l=1}^{M}g_l(\partial/\partial x_l)\right)e^{i\mathbf{k}\cdot\mathbf{x}} =-\sigma^\alpha|\mathbf{g}\cdot\mathbf{k}|^\alpha e^{i\mathbf{k}\cdot\mathbf{x}}$.

With the characteristic function
\begin{equation}
F_\mathbf{x}(\mathbf{k},t)\equiv\langle{e^{i\mathbf{k}\cdot\mathbf{x}}}\rangle
=\int w(\mathbf{x},t)\,e^{i\mathbf{k}\cdot\mathbf{x}}\mathrm{d}^M\mathbf{x}\;,
\label{eq106}
\end{equation}
one can write the Fourier transform
\begin{equation}
w(\mathbf{x},t)=\frac{1}{(2\pi)^{M}}\int F_\mathbf{x}(\mathbf{k},t)\,e^{-i\mathbf{k}\cdot\mathbf{x}}\mathrm{d}^M\mathbf{k}\;,
\label{eq107}
\end{equation}
and
\begin{equation}
\dot\Phi_t^{(\xi)}(i\hat{Q})w(\mathbf{x},t) =\frac{-\sigma^\alpha}{(2\pi)^M}\int|\mathbf{g}\cdot\mathbf{k}|^\alpha  F_\mathbf{x}(\mathbf{k},t)\,e^{-i\mathbf{k}\cdot\mathbf{x}}\mathrm{d}^M\mathbf{k}\,.
\label{eq108}
\end{equation}

\subsection{Populations of quadratic integrate-and-fire neurons}
\label{ssec:popQIF}
Our consideration in Secs.~\ref{ssec:DCnG} and \ref{ssec:fracFP} was performed in the general form. Further advance cannot be made with the same level of generality; it must be specific for a specific form of the deterministic part of Eq.~(\ref{eq001}). Henceforth, we consider populations of quadratic integrate-and-fire neurons (QIFs).~\cite{Izhikevich-2007}

Let us consider the recursive network of QIFs with global synaptic coupling and endogenic noise:
\begin{align}
&\dot{V}_j=V_j^2+I_j\;,
\label{eq110}
\\
&I_j=\eta_j+\sigma\xi_j(t)+Js(t)+I(t)\;,
\label{eq111}
\end{align}
where $V_j$: the membrane voltage; $\eta_j$: the excitability parameter of individual neuron, an isolated neuron is excitable for $\eta_j<0$ and periodically spiking otherwise; $I(t)$: the external input current; independent $\alpha$-stable endogenic ({\it or} intrinsic) noises $\xi_j(t)$ are normalized [see Eqs.~(\ref{eq001}), (\ref{eq003}) for amplitude $c=1$].
When $V_j$ reaches the threshold value $B_{+}$ it is reset to $-B_{-}$ and a synaptic spike is generated.~\cite{Ermentrout-Kopell-1986} The values of $B_{\pm}$ are large and the theoretical results converge for $B_{\pm}\to\infty$.
In the limit $B_{\pm}\to\infty$, the inequality between $B_+$ and $B_-$ is found to effect the macroscopic dynamics of the population in the presence of the electrical gap coupling between neurons,~\cite{Montbrio-Pazo-2020} which is not considered in the model~(\ref{eq110},\ref{eq111}); its effect is controlled by the ratio $B_+/B_-$.
The impact of inequality of finite $B_\pm$ on the macroscopic dynamics without gap junctions was also thoroughly studied.~\cite{Gast-Solla-Kennedy-2023}
Specifically for the model~(\ref{eq110},\ref{eq111}), without the loss of generality, one can consider $B_+/B_-=1$ and $B_{\pm}\to\infty$.
The input synaptic current from other neurons $Js(t)$ is characterised by the coupling coefficient $J$, which is negative for an inhibitory coupling, and a common field
\[
s(t)=\frac{1}{N}\sum_{j=1}^{N}\sum_{n}\delta(t-t_j^{(n)})\;,
\]
where $N$ is the number of neurons and $t_j^{(n)}$ is the time instant of the $n$th firing event of the $j$th neuron. In the thermodynamic limit of a large population, $N\to\infty$, the common field $s(t)=r(t)$, where the firing rate $r(t)$ is the probability rate of the firing event of an individual neuron averaged over the population.

We consider a heterogeneous population with $\eta_j$ distributed according to a Cauchy distribution
\begin{equation}
g(\eta)=\frac{\pi^{-1}\Delta}{(\eta-\eta_0)^2+\Delta^2}\;,
\label{eq112}
\end{equation}
where $\eta_0$ is the median of the distribution and $\Delta$ is the half-width at half-maximum (HWHM).

Let us index the QIFs by the value of parameter $\eta_j$.
For (\ref{eq110},\ref{eq111}), fractional Fokker--Planck equation~(\ref{eq105}) with one-dimensional $\mathbf{x}=V$ for the $\eta$-subpopulation reads
\begin{equation}
\frac{\partial w_\eta}{\partial t}+
\frac{\partial}{\partial V}\Big((I_\eta+V^2)w_\eta\Big)
-\dot\Phi_t^{(\xi)}\Big(i\sigma\frac{\partial}{\partial V}\Big)w_\eta=0\,,
\label{eq113}
\end{equation}
with
\[
I_\eta=\eta+Jr(t)+I(t)\,.
\]
Here the population-mean firing rate
\begin{equation}
r(t)=\int_{-\infty}^{+\infty}r_\eta(t)\,g(\eta)\,\mathrm{d}\eta\;,
\label{eq114}
\end{equation}
where the $\eta$-subpopulation firing rate $r_\eta$ is the probability density flux $q_\eta(V,t)$ at $V=\pm\infty$.

The probability density flux
\begin{align}
q_\eta(V,t)&=(I_\eta+V^2)\,w_\eta(V,t)
\nonumber\\
&\quad
-\left(\frac{\partial}{\partial V}\right)^{-1}\dot\Phi_t^{(\xi)}\Big(i\sigma\frac{\partial}{\partial V}\Big)w_\eta(V,t)\;,
\nonumber
\end{align}
where $(\partial/\partial V)^{-1}f(V)\equiv\int_{-\infty}^V\mathrm{d}V_1f(V_1)$, the diffusive (latter) term vanishes at $V=\pm\infty$ by virtue of its physical origin as $w_\eta(V,t)$ and all its derivatives vanish at $\pm\infty$ for all $\alpha>0$. Therefore, the relation
\begin{equation}
r_\eta(t)=q_\eta|_{V=\infty}=\lim_{V\to\infty}V^2w_\eta(V,t)
\label{eq115}
\end{equation}
holds for all $\alpha>0$.

\subsection{Logarithm of characteristic function and pseudocumulants}
\label{ssec:CF&PC}
\paragraph{Characteristic function.}
One can project fractional Fokker--Planck equation~(\ref{eq113}) into the Fourier space and find the evolution equation for the characteristic function $\mathcal{F}_\eta(k,t)=\langle{e^{ikV_\eta}}\rangle$ of the $\eta$-subpopulation (see Ref.~\citen{Goldobin-Volo-Torcini-2021} for a rigorous derivation);
\begin{equation}
\frac{\partial \mathcal{F}_\eta}{\partial t}
=ik\left[I_\eta\mathcal{F}_\eta-\frac{\partial^2}{\partial k^2}\mathcal{F}_\eta\right]
-\sigma^\alpha|k|^\alpha\mathcal{F}_\eta\;.
\label{eq116}
\end{equation}
With distribution~(\ref{eq112}), one can employ the residue theorem and find the evolution equation for the population-mean characteristic function~\cite{Goldobin-Volo-Torcini-2021}
\[
F(k,t)=\int_{-\infty}^{+\infty}g(\eta)\,\mathcal{F}_\eta(k,t)\,\mathrm{d}\eta\,:
\]
\begin{equation}
\frac{\partial F}{\partial t}
=ik\left[I_0F-\frac{\partial^2F}{\partial k^2}\right]
 -|k|\Delta F -\sigma^\alpha|k|^\alpha F\;,
\label{eq117}
\end{equation}
where $I_0=\eta_0+Jr(t)+I(t)$.

For the logarithm of characteristic function, $\Phi(k,t)=\ln{F(k,t)}$, one finds
\begin{equation}
\frac{\partial\Phi}{\partial t}
=ik\left[I_0-\frac{\partial^2\Phi}{\partial k^2}
 -\left(\frac{\partial\Phi}{\partial k}\right)^2\right]
 -\Delta|k| -\sigma^\alpha|k|^\alpha\,.
\label{eq118}
\end{equation}
From definition~(\ref{eq106}), one can see the symmetry properties of $\Phi(\mathbf{k},t)$ [and $F(\mathbf{k},t)$]:
\begin{equation}
\Phi^\ast(\mathbf{k},t)=\Phi(-\mathbf{k},t)\;,
\label{eq119}
\end{equation}
where the asterisk indicates complex conjugate.

For $\alpha=1$ and $2$, Eq.~(\ref{eq118}) and the symmetry property~(\ref{eq119}) suggest the most general form of the integer power-series expansion for $\Phi(k,t)$:
\begin{align}
\Phi(k,t)=-a(t)|k|+iv(t)k-\frac{q_2(t)|k|^2+ip_2(t)|k|k}{2}
\nonumber\\
-\frac{q_3(t)|k|^3+ip_3(t)|k|^2k}{3}-\dots \qquad\qquad
\nonumber\\
-\frac{q_m(t)|k|^m+ip_m(t)|k|^{m-1}k}{m}-\dots\;,
\nonumber
\end{align}
where the series starts from the piecewise linear term as $\Phi(0,t)=\ln{F(0,t)}=0$ by construction; for this series, one can alternatively write
\begin{align}
&\Phi(k>0)=-\sum_{m=1}^{\infty}W_m\frac{k^m}{m}\;,
\label{eq122}\\
&W_1=a-iv\;,\quad
W_m=q_m+ip_m\;.
\nonumber
\end{align}
Recently,\cite{Goldobin-Volo-Torcini-2021} the framework of quantities $W_m$ was successfully employed for the theoretical study of noise-induced collective dynamics of populations of QIFs and $W_m$ received the name `pseudocumulants.'

For a noninteger rational $\alpha=L/N$, where $L$ and $N$ are natural numbers, Eq.~(\ref{eq118}) does not admit integer power series but might admit
\begin{align}
&\Phi(k>0)=-\sum_{m=N}^{\infty}W_\frac{m}{N}\frac{k^\frac{m}{N}}{m/N}\;,
\label{eq122N}\\
&\Phi(k<0)=-\sum_{m=N}^{\infty}W_\frac{m}{N}^\ast\frac{|k|^\frac{m}{N}}{m/N}\;,
\nonumber
\end{align}
with nonzero fractional power terms, and the elements with $m<N$ are inadmissible as they induce the elements with negative $m$ in Eq.~(\ref{eq118}).

For an irrational $\alpha$, Eq.~(\ref{eq118}) cannot admit anything but ``mixed'' power series:
\begin{align}
&\Phi(k>0)=-\sum_{m,n}W_{m,n}\frac{k^{m+n\alpha}}{m+n\alpha}\;,
\label{eq122mn}
\end{align}
where $W_{m,n}=0$ for $m+n\alpha<1$.

\paragraph{Boundary conditions for $\Phi(k,t)$.} Partial differential equation~(\ref{eq118}) is a second order equation with respect to $k$, and generally one must have two boundary conditions for $\Phi(k,t)$. First, by definition, $F(0,t)=1$ and thus
\begin{equation}
\Phi(0,t)=0\;.
\label{eq120}
\end{equation}
Second, for a physically meaningful distribution $w(V,t)$, its inverse Fourier transform $F(k,t)$ must tend to zero at $k\to\infty$. Since $F=e^\Phi$, one must claim
\begin{equation}
\lim_{k\to\pm\infty}\mathrm{Re}[\Phi(k,t)]=-\infty\;.
\label{eq121}
\end{equation}

\paragraph{Pseudocumulants.}
In noise-free heterogeneous populations, $\sigma=0$ and $\Delta>0$, a Cauchy distribution of voltage $V$ was found to be attracting.~\cite{Montbrio-Pazo-Roxin-2015,Pietras-etal-2023}~\footnote{As shown in Ref.~\cite{Montbrio-Pazo-Roxin-2015}, a Cauchy distribution of $V_j$ in system~(\ref{eq110})--(\ref{eq111}) with $\sigma=0$ is equivalent to a wrapped Cauchy distribution~\cite{Ley-Verdebout-2017} of phases $\phi_j=2\arctan{V_j}$, which corresponds to the Ott--Antonsen ansatz. Thus, the problem of the attractivity of this distribution turns into the problem of attractivity of the Ott--Antonsen manifold~\cite{Ott-Antonsen-2008} for the case of nonidentical oscillators~\cite{Ott-Antonsen-2009,Mirollo-2012,Pietras-Daffertshofer-2016}} A Cauchy distribution
\[
w_\mathrm{LD}(V,t)=\frac{\pi^{-1} a(t)}{a(t)^2+[V-v(t)]^2}
\]
corresponds to the characteristic function $F_\mathrm{LD}(k)=\exp[-a|k|+ivk]$, and
\[
\Phi_\mathrm{LD}(k)=-a|k|+ivk\;.
\]
Indeed, Eq.~(\ref{eq118}) with $\sigma=0$ admits the latter solution.
The distribution width $a$ determines the firing rate:
\[
r_\mathrm{LD}(t)=\lim_{V\to\infty}V^2w_\mathrm{LD}(V,t)=\frac{a(t)}{\pi}\;.
\]

The framework of pseudocumulant expansion~(\ref{eq122}) was introduced for dealing with perturbations of a Cauchy distribution.~\cite{Goldobin-Volo-Torcini-2021} This approach was employed to study the effect of a Gaussian noise ($\alpha=2$) in populations of QIFs. The states of the form $W_m=W_1\delta_{1m}$, where $\delta_{nm}$ is the Kronecker delta, correspond to a Cauchy distribution and higher pseudocumulants $W_{m>1}$ describe deviations from it.

After substitution~(\ref{eq122}), partial differential equation~(\ref{eq118}) with $\alpha=1$ (Cauchy noise) or $2$ (Gaussian noise) yields an infinite equation chain for pseudocumulants $W_m$\,;
\begin{align}
&\dot{W}_m=(\Delta-iI_0)\delta_{1m}+\alpha\sigma^\alpha\delta_{\alpha m}
\nonumber\\
&\qquad{} +im\Big(-mW_{m+1}+\sum\nolimits_{n=1}^{m}W_nW_{m+1-n}\Big)\,,
\label{eq123}
\\
&\qquad\mbox{ with }\quad
 r(t)=\frac{1}{\pi}\mathrm{Re}(W_1)\;.
\nonumber
\end{align}
The convergence of a pseudocumulant series allows one to use finite truncations of the infinite equation chain~(\ref{eq123}), where one sets $W_m=0$ for $m$ larger than some truncation threshold $m_\mathrm{tr}$. Such truncations yield low dimensional neural mass models.

However, substitution~(\ref{eq122}) is questionable for noninteger values of $\alpha$ as $|k|^\alpha$ with $0<\alpha<1$ or $1<\alpha<2$ cannot be adequately represented by a series in integer powers of $k$. Below, we consider partial differential equation~(\ref{eq118}) for noninteger $\alpha$.
One of the aims of this paper is also to examine the options~(\ref{eq122N}) and (\ref{eq122mn}) for the generalization of the concept of pseudocumulants to the fractional $\alpha$ case.

\section{Noninteger $\alpha$}
\label{sec:fracA}
\subsection{Linear approximation for a time-independent macroscopic state}
Let us consider the case of a weak noise and use $\varepsilon=\sigma^\alpha$ as a smallness parameter. We substitute
\[
\Phi=\Phi_0+\varepsilon\Phi_1+\dots
\]
into Eq.~(\ref{eq118}) for $k>0$ [the case of $k<0$ is similar according to the symmetry~(\ref{eq119})] and obtain the following:
\\
$\bullet$~In the $\varepsilon^0$-order,
\begin{equation}
\Phi_0=-W_{1,0}^{(0)}k\;,
\label{eq201}
\end{equation}
\begin{equation}
\frac{\mathrm{d}}{\mathrm{d}t}W_{1,0}^{(0)}
=\Delta-iI_0+i\big(W_{1,0}^{(0)}\big)^2\;.
\label{eq202}
\end{equation}
Here the superscript $(0)$ indicates the leading order solution for $\varepsilon=0$.
Eq.~(\ref{eq202}) admits two time-independent solutions $W_{1,0}^{(0)}=\pm\sqrt{I_0+i\Delta}$. Since $I_0$ can be both positive and negative and $\Delta>0$, the plus-solution always belongs to the first quadrant of the complex plane, while the minus-solution does to the third one. Hence, only
\begin{equation}
W_{1,0}^{(0)}=+\sqrt{I_0+i\Delta}
\label{eq203}
\end{equation}
satisfies the condition~(\ref{eq121}) [see Eq.~(\ref{eq201})].
\\
$\bullet$~In the $\varepsilon^1$-order, a time-independent solution obeys
\begin{equation}
\Phi_1^{\prime\prime}-2W_{1,0}^{(0)}\Phi_1^\prime-ik^{\alpha-1}=0\;,
\label{eq204}
\end{equation}
where the prime denotes the $k$-derivative.

The formal solution of Eq.~(\ref{eq204}) reads
\begin{equation}
\Phi_1^\prime(k>0)
=e^{2W_{1,0}^{(0)}k}\bigg(C_0 +i\int\limits_{0}^{k}\mathrm{d}k_1k_1^{\alpha-1}e^{-2W_{1,0}^{(0)}k_1}\bigg)\,,
\label{eq205}
\end{equation}
where the integration constant $C_0$ is determined by condition~(\ref{eq121}). For $\Delta>0$, $W_{1,0}^{(0)}$ has a positive real part and a nonzero imaginary one,  i.e., $\exp[2W_{1,0}^{(0)}k]$ grows and oscillates with $k$; therefore, condition~(\ref{eq121}) can be satisfied only if
$C_0 +i\int_{0}^{+\infty}\mathrm{d}k_1k_1^{\alpha-1}e^{-2W_{1,0}^{(0)}k_1}=0$.
Hence,~\footnote{For $W_{1,0}^{(0)}=\sqrt{I_0+i\Delta}$ laying in the first quadrant of the complex plane, one can calculate the integral $\int_{0}^{+\infty}\mathrm{d}k_1k_1^{\alpha-1}e^{-2W_{1,0}^{(0)}k_1}
 =\big(2W_{1,0}^{(0)}\big)^{-\alpha}\int_0^{+\infty\cdot W_{1,0}^{(0)}} \mathrm{d}y\,y^{\alpha-1}e^{-y}
 =\big(2W_{1,0}^{(0)}\big)^{-\alpha}\Big[
  \int_0^{+\infty}\mathrm{d}y\,y^{\alpha-1}e^{-y}
 +\int_{+\infty}^{+\infty\cdot W_{1,0}^{(0)}}\mathrm{d}y\,y^{\alpha-1}e^{-y}\Big] =\big(2W_{1,0}^{(0)}\big)^{-\alpha}\big[\Gamma(\alpha)+0\big]$.}
\begin{equation}
C_0=-i\frac{\Gamma(\alpha)}{\big(2W_{1,0}^{(0)}\big)^\alpha}\;,
\label{eq206}
\end{equation}
where $\Gamma(\cdot)$ is the complete gamma function.
Therefore, Eq.~(\ref{eq205}) can be rewritten as
\[
\Phi_1^\prime(k>0)
=-i\int\limits_{k}^{+\infty}\mathrm{d}k_1k_1^{\alpha-1}e^{2W_{1,0}^{(0)}(k-k_1)}\;,
\]
and, after integration with condition $\Phi_1(0)=0$ (\ref{eq120}), one obtains
\begin{equation}
\Phi_1(k>0)
=-i\int\limits_{0}^{k}\mathrm{d}k_1 \int\limits_{k_1}^{+\infty}\mathrm{d}k_2k_2^{\alpha-1}e^{2W_{1,0}^{(0)}(k_1-k_2)}\;.
\label{eq207}
\end{equation}

Now we must check that condition~(\ref{eq206}) is sufficient for solution~(\ref{eq207}) to satisfy (\ref{eq121}). For $k\gg 1$, the integral
\begin{align}
&\int\limits_{k_1}^{+\infty}\mathrm{d}k_2k_2^{\alpha-1}e^{2W_{1,0}^{(0)}(k_1-k_2)}
\nonumber\\
&\qquad{}
\approx k_1^{\alpha-1}\int\limits_0^{+\infty}\mathrm{d}k_2e^{-2W_{1,0}^{(0)}k_2}=\frac{k_1^{\alpha-1}}{2W_{1,0}^{(0)}}\;,
\nonumber
\end{align}
and
\[
-i\int\limits_0^k\mathrm{d}k_1\frac{k_1}{2W_{1,0}^{(0)}}
 =\frac{-ik^\alpha}{2\alpha W_{1,0}^{(0)}}
 =\frac{k^\alpha(-W_{1,0\mathrm{i}}^{(0)}-iW_{1,0\mathrm{r}}^{(0)})}{2\alpha\big|W_{1,0}^{(0)}\big|^2}\;,
\]
where $W_{1,0}^{(0)}\equiv W_{1,0\mathrm{r}}^{(0)}+iW_{1,0\mathrm{i}}^{(0)}$ and, for Eq.~(\ref{eq203}), one can calculate $W_{1,0\mathrm{i}}^{(0)}=\big[(\sqrt{I_0^2+\Delta^2}-I_0)/2\big]^{1/2}$\,. Hence, $\mathrm{Re}\,\Phi_1|_{k=\pm\infty}=-\infty$ and solution~(\ref{eq207}) satisfies the boundary conditions.

Summarizing,
\begin{align}
\Phi(k>0)&=-W_{1,0}^{(0)}k -i\sigma^\alpha\int\limits_{0}^{k}\mathrm{d}k_1 \int\limits_{k_1}^{+\infty}\mathrm{d}k_2\,k_2^{\alpha-1}e^{2W_{1,0}^{(0)}(k_1-k_2)}
\nonumber\\
&\qquad\qquad\quad{}
 +\mathcal{O}\left([\sigma^\alpha]^2\right)\;.
\label{eq208}
\end{align}

\subsection{Expansion of $\Phi(k)$}
In order to put solution~(\ref{eq208}) into the context of the pseudocumulant expansion, we consider the series of (\ref{eq208}) for small $k$. For this task, it is convenient to recast the integral $\int_{k_1}^{+\infty}\dots\mathrm{d}k_2$ into the form with the integration from $0$, i.e., $\int_0^{k_1}\dots\mathrm{d}k_2$ [see~(\ref{eq205})];
\begin{align}
\Phi(k>0)&=-W_{1,0}^{(0)}k +\sigma^\alpha\int\limits_{0}^{k}\mathrm{d}k_1
e^{2W_{1,0}^{(0)}k_1}
\nonumber\\
&
\times\bigg(C_0+i\int\limits_{0}^{k_1}\mathrm{d}k_2k_2^{\alpha-1}e^{-2W_{1,0}^{(0)}k_2}\bigg)
+\mathcal{O}\left([\sigma^\alpha]^2\right)\;.
\nonumber
\end{align}
In the latter equation, we can substitute expansions of all functions in series of $k$ and find
\begin{align}
\Phi(k>0)&=-W_{1,0}^{(0)}k +\sigma^\alpha\bigg[C_0\bigg(k+W_{1,0}^{(0)}k^2
+\frac23\big[W_{1,0}^{(0)}\big]^2k^3
 \bigg)
\nonumber\\
&\qquad{}
+\frac{ik^{\alpha+1}}{\alpha(\alpha+1)} +\frac{i2W_{1,0}^{(0)}k^{\alpha+2}}{\alpha(\alpha+1)(\alpha+2)}
 \bigg]
\nonumber\\[7pt]
&\qquad\qquad{}
 +\mathcal{O}\left(\sigma^\alpha k^4,\sigma^\alpha k^{\alpha+3},[\sigma^\alpha]^2\right)\;.
\nonumber
\end{align}
Substituting $C_0$ (\ref{eq206}), we obtain
\begin{align}
&\Phi(k>0)=-W_{1,0}^{(0)}k +\frac{\sigma^\alpha}{2^\alpha}\left[-\frac{i\Gamma(\alpha)\,k}{\big[W_{1,0}^{(0)}\big]^\alpha} -\frac{i\Gamma(\alpha)\,k^2}{\big[W_{1,0}^{(0)}\big]^{\alpha-1}}
\right.
\nonumber\\
&\qquad\left.
-\frac{i2\Gamma(\alpha)\,k^3}{3\big[W_{1,0}^{(0)}\big]^{\alpha-2}}
+\frac{i2^\alpha k^{\alpha+1}}{\alpha(\alpha+1)} +\frac{i2^{\alpha+1}W_{1,0}^{(0)}k^{\alpha+2}}{\alpha(\alpha+1)(\alpha+2)}
 \right]
\nonumber\\[10pt]
&\qquad\qquad\quad{}
 +\mathcal{O}\left(\sigma^\alpha k^4,\sigma^\alpha k^{\alpha+3},[\sigma^\alpha]^2\right)\;.
\label{eq209}
\end{align}

One can write $\Phi(k)$ with the leading terms of expansion~(\ref{eq122mn}) for both positive and negative $k$:
\begin{align}
\Phi(k)&=-W_{1,0\mathrm{r}}|k|-iW_{1,0\mathrm{i}}k -W_{2,0\mathrm{r}}\frac{k^2}{2}-iW_{2,0\mathrm{i}}\frac{k|k|}{2}
\nonumber\\
&\qquad{}
-W_{1,1\mathrm{r}}\frac{|k|^{\alpha+1}}{\alpha+1}-iW_{1,1\mathrm{i}}\frac{|k|^\alpha k}{\alpha+1}-\dots\,,
\label{eq210}
\end{align}
where $W_{m,n\mathrm{r}}$ and $W_{m,n\mathrm{i}}$ are the real and imaginary parts of $W_{m,n}$ 
and, according to~(\ref{eq209}) and (\ref{eq203}),
\begin{align}
W_{1,0}&=W_{1,0}^{(0)}+\frac{i\sigma^\alpha\Gamma(\alpha)}{\big[2W_{1,0}^{(0)}\big]^\alpha}
 +\mathcal{O}\left([\sigma^\alpha]^2\right)
\label{eq211}
\\
&=\sqrt{\frac{\sqrt{I_0^2+\Delta^2}+I_0}{2}}+i\sqrt{\frac{\sqrt{I_0^2+\Delta^2}-I_0}{2}}
\nonumber\\
&\quad
{}
+\frac{\sigma^\alpha\Gamma(\alpha)}{2^\alpha(I_0^2+\Delta^2)^\frac{\alpha}{2}}\bigg[
\sin\bigg(\frac{\alpha}{2}\arccos\frac{I_0}{\sqrt{I_0^2+\Delta^2}}\bigg)
\nonumber\\
&\qquad
{}
+i\cos\bigg(\frac{\alpha}{2}\arccos\frac{I_0}{\sqrt{I_0^2+\Delta^2}}\bigg)
\bigg]
+\mathcal{O}\left([\sigma^\alpha]^2\right)\,.
\label{eq212}
\end{align}

\subsection{Probability distribution}
With~(\ref{eq210}), one can find (see Appendix~\ref{app:PDF})
\begin{align}
&w(V)=
\bigg(1+\frac{W_{2,0\mathrm{r}}}{2}\frac{\partial^2}{\partial V^2}
 +\frac{W_{2,0\mathrm{i}}}{2}\frac{\partial}{\partial V}\left|\frac{\partial}{\partial W_{1,0\mathrm{r}}}\right|
\nonumber\\
&\qquad{}
 +\frac{W_{1,1\mathrm{r}}}{\alpha+1}\frac{\partial^2}{\partial V^2}\left|\frac{\partial}{\partial W_{1,0\mathrm{r}}}\right|^{\alpha-1}
 +\frac{W_{1,1\mathrm{i}}}{\alpha+1}\frac{\partial}{\partial V}\left|\frac{\partial}{\partial W_{1,0\mathrm{r}}}\right|^{\alpha}
\nonumber\\
&\qquad\qquad
 +\dots\bigg)
 \frac{\pi^{-1}W_{1,0\mathrm{r}}}{W_{1,0\mathrm{r}}^2+(V+W_{1,0\mathrm{i}})^2}\;.
\label{eq213}
\end{align}
Here $W_{1,0}$ is given by Eq.~(\ref{eq211}) [or (\ref{eq212})].

In Appendix~\ref{app:PDF} we show that with distribution~(\ref{eq213}), one can calculate firing rate~(\ref{eq115}):
\begin{equation}
r=\frac{\mathrm{Re}(W_{1,0})}{\pi}\;,
\label{eq214}
\end{equation}
and this formula is valid with any number of corrections $W_{m,n}$. Substituting~(\ref{eq212}), we find
\begin{align}
r&=\frac{\sqrt{\sqrt{I_0^2+\Delta^2}+I_0}}{\sqrt{2}\pi}
\nonumber\\
&
{}
+\frac{\sigma^\alpha\Gamma(\alpha) \sin\!\Big(\frac{\alpha}{2}\arccos\frac{I_0}{\sqrt{I_0^2+\Delta^2}}\Big)} {2^\alpha\pi(I_0^2+\Delta^2)^\frac{\alpha}{2}}
+\mathcal{O}\left([\sigma^\alpha]^2\right)\,.
\label{eq215}
\end{align}
Further, the mean value of $w(V)$ is $-W_{1,0\mathrm{i}}$ and not affected by the higher order corrections $W_{m,n}$ (Appendix~\ref{app:PDF}):
\begin{align}
&\langle{V}\rangle
=-\mathrm{Im}(W_{1,0})\;.
\label{eq216}
\end{align}
Substituting~(\ref{eq212}), we find
\begin{align}
\langle{V}\rangle &=-\sqrt{\frac{\sqrt{I_0^2+\Delta^2}-I_0}{2}}
\nonumber\\
&
{}
-\frac{\sigma^\alpha\Gamma(\alpha) \cos\!\Big(\frac{\alpha}{2}\arccos\frac{I_0}{\sqrt{I_0^2+\Delta^2}}\Big)} {2^\alpha(I_0^2+\Delta^2)^\frac{\alpha}{2}}
+\mathcal{O}\left([\sigma^\alpha]^2\right)\,.
\label{eq217}
\end{align}
Equations~(\ref{eq215}) and (\ref{eq217}) constitute the result~(\ref{eq212}) of the linear approximation for time-independent regimes in terms of physically meaningful macroscopic observables.

\section{Macroscopic states of homogeneous populations of QIFs subject to $\alpha$-stable noise}
\label{sec:QIFs}
In this section we employ the results for firing rate~(\ref{eq215}) and mean voltage~(\ref{eq217}) to construct a self-consistent mathematical description of the macroscopic states of the population of QIFs with global synaptic coupling (\ref{eq110},\ref{eq111}). Here $I_0=\eta_0+Jr$. In the studies with circular and pseudo- cumulants for a Gaussian noise,~\cite{Goldobin-Dolmatova-2020,Goldobin-2021,Goldobin-Volo-Torcini-2021} the  most challenging cases were that of a small or vanishing heterogeneity $\Delta$. Therefore, it will be instructive to demonstrate the application of our theoretical findings to the case of $\Delta=0$, where in the absence of noise the OA (and MPR) manifold is marginally stable. The results for a less technically challenging case of heterogeneous populations will be provided in the next section.

One can distinguish two kinds of macroscopic regimes. (i)~For $I_0>0$, the macroscopic states are mean-field driven ones, where the individual QIFs are forced above their excitability threshold. (ii)~For $I_0<0$, the macroscopic states are noise-driven ones, where the individual QIFs are excitable and all the firing events are induced by noise fluctuations.

\subsection{Mean-field driven regimes}
For $I_0>0$ and $\Delta=0$, Eq.~(\ref{eq215}) takes the form
\begin{equation}
r=\frac{\sqrt{I_0}}{\pi}
+\mathcal{O}(\sigma^{2\alpha})\;.
\label{eq301}
\end{equation}
With $I_0=\eta_0+Jr$, neglecting the $\sigma^{2\alpha}$-corrections, one finds the self-consistency condition
\begin{equation}
r=\frac{J\pm\sqrt{J^2+4\pi^2\eta_0}}{2\pi^2}\;;
\label{eq302}
\end{equation}
a parametric form of the dependence $r$ versus $\eta_0$ will be also useful
\begin{equation}
\eta_0=-Jr+\pi^2r^2\;.
\label{eq303}
\end{equation}
With (\ref{eq303}) one can find the saddle-node bifurcation point from the condition $\mathrm{d}\eta_0/\mathrm{d}r=0$\,:
\begin{equation}
\eta_{0\mathrm{mf}}=-\frac{J^2}{4\pi^2}\;,
\qquad
r_\mathrm{mf}=\frac{J}{2\pi^2}\;.
\label{eq304}
\end{equation}

With Eq.~(\ref{eq217}), one obtains for these regimes a noise-induced negative shift of mean voltage:
\begin{equation}
\langle{V}\rangle
=-\frac{\sigma^\alpha\Gamma(\alpha)} {2^\alpha(\eta_0+Jr)^\alpha}
+\mathcal{O}\left(\sigma^{2\alpha}\right)\;.
\label{eq305}
\end{equation}

\subsection{Noise-driven regimes}
For $I_0<0$ and $\Delta=0$, Eq.~(\ref{eq215}) takes the form
\begin{equation}
r=\frac{\Gamma(\alpha)}{\pi}\left(\frac{\sigma}{2|I_0|}\right)^\alpha\sin\frac{\alpha\pi}{2}
+\mathcal{O}(\sigma^{2\alpha})\;,
\label{eq306}
\end{equation}
where the firing rate is purely noise-induced.
With $I_0=\eta_0+Jr<0$, neglecting the $\sigma^{2\alpha}$-corrections, one can write the self-consistency condition as
\begin{equation}
\eta_0=-Jr -\frac{\sigma}{2}
 \left(\frac{\Gamma(\alpha)}{\pi r}\sin\frac{\alpha\pi}{2}\right)^\frac{1}{\alpha}\;,
\label{eq307}
\end{equation}
which gives a parametric form of the dependence $r$ versus $\eta_0$.
With (\ref{eq306}) one can find the saddle--node bifurcation point from the condition $\mathrm{d}\eta_0/\mathrm{d}r=0$\,:
\begin{align}
\eta_{0\mathrm{nd}}&=-(1+\alpha) \left(\frac{\sigma^\alpha\Gamma(\alpha) J}{\pi(2\alpha)^\alpha} \sin\frac{\alpha\pi}{2}\right)^\frac{1}{1+\alpha},
\label{eq308}
\\
r_\mathrm{nd}&=\left(\frac{\sigma^\alpha\Gamma(\alpha)}{\pi(2\alpha J)^\alpha} \sin\frac{\alpha\pi}{2}\right)^\frac{1}{1+\alpha}.
\label{eq309}
\end{align}

With Eq.~(\ref{eq217}), one obtains for these regimes a noise-induced shift of mean voltage
\begin{align}
\langle{V}\rangle
=-\sqrt{|\eta_0+Jr|} -\frac{\sigma^\alpha\Gamma(\alpha)\cos\frac{\alpha\pi}{2}}{2^\alpha|\eta_0+Jr|^\alpha}
+\mathcal{O}\left(\sigma^{2\alpha}\right)\;;
\label{eq310}
\end{align}
the shift is positive for $\alpha>1$ and negative for $\alpha<1$.

\begin{figure}[!t]
\includegraphics[width=0.45\textwidth]%
 {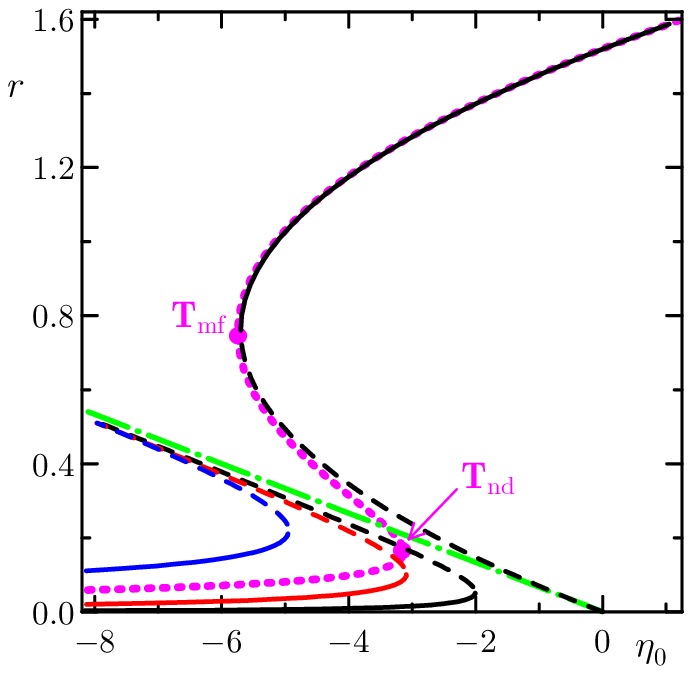}
\caption{
For a homogeneous population of quadratic integrate-and-fire neurons~(\ref{eq110},\ref{eq111}) subject to $\alpha$-stable noise.
The exact solution for a Cauchy noise (\ref{eq313}) is plotted with the bold magenta dotted curve;
the dependence of the firing rate $r$ versus $\eta_0$ exhibits multistability between the mean-field and noise driven macroscopic regimes (the upper and lower branches of solution).
Points $\mathbf{T}_\mathrm{mf}$ and $\mathbf{T}_\mathrm{nd}$ mark the saddle--node bifurcation points of mean-field and noise driven regimes, respectively. The green dash-dotted line depicts the states with $I_0=\eta_0+Jr=0$. Above $I_0=0$, the black curves represent the approximate mean-field driven solution~(\ref{eq302}), for which the firing rate $r$ is independent of $\alpha$, to the leading order. The solid curve is a stable solution. Below $I_0=0$, the approximate noise driven solution~(\ref{eq307}) is plotted for $\alpha=1/2$, $1$, $3/2$ (from left to right: blue, red, black); the solid curves are stable solutions. Parameters: $J=15$ and $\sigma=1$.
}
  \label{fig1}
\end{figure}

For both mean-field and noise driven regimes, the approximate solutions near $I_0=\eta_0+Jr=0$ (dash-dotted line in Fig.~\ref{fig1}) are inaccurate, since for $I_0=0$ in (\ref{eq301}), (\ref{eq305}), (\ref{eq306}), and (\ref{eq310}) the leading order vanishes or diverges. The $\sigma^{2\alpha}$- and higher terms become non-negligible. However, these inaccurate branches of solutions are unstable (see subsection~\ref{ssec:QIFstab}), while the stable branches are farther from this line and reasonably accurate (see Fig.~\ref{fig1}).

The hysteretic transition between the mean-field and noise driven regimes occurring in the parameter domain where they overlap was confirmed with the direct numerical simulations of the microscopic dynamics of a homogeneous population of $1000$ QIFs (\ref{eq110},\ref{eq111}); the results are presented in Fig.~\ref{fig2}. See Appendix~\ref{app:microDNS} for details on the direct numerical simulation of the microscopic dynamics.

\subsection{The exactly solvable case of a Cauchy noise}
The case of Cauchy noise ($\alpha=1$) can be solved exactly:~\cite{Pietras-etal-2023,Pyragas2-2023,Clusella-Montbrio-2022} in Eq.~(\ref{eq118}), $\sigma^\alpha$- and $\Delta$-terms can be combined and yield an effective disorder parameter $\Delta+\sigma\to\Delta_\sigma$. For this case, one can take Eqs.~(\ref{eq215}) and (\ref{eq217}), set $\sigma^\alpha$ to zero and then replace $\Delta$ with $\Delta+\sigma$. For a homogeneous population, one finds
\begin{align}
r&=\frac{\sqrt{\sqrt{I_0^2+\sigma^2}+I_0}}{\sqrt{2}\pi}\;,
\label{eq311}
\\
\langle{V}\rangle &=-\sqrt{\frac{\sqrt{I_0^2+\sigma^2}-I_0}{2}}\;.
\label{eq312}
\end{align}
From Eq.~(\ref{eq311}), one can derive a parametric form of the dependence $r$ versus $\eta_0$:
\begin{equation}
\eta_0=-Jr +\pi^2r^2-\frac{\sigma^2}{4\pi^2r^2}\;.
\label{eq313}
\end{equation}
The saddle--node bifurcation points can be obtained from the condition $\mathrm{d}\eta_0/\mathrm{d}r=0$ (\ref{eq313}) in a parametric form, parameterized by $r_\ast$
(earlier reported in Appendix~C of Ref.~\citen{Pietras-Devalle-Roxin-etal-2019}):
\begin{align}
J_\ast&=2\pi^2r_\ast+\frac{\sigma^2}{2\pi^2r_\ast^3}\;,
\label{eq314}
\\
\eta_{0\ast}&=-\pi^2r_\ast^2-\frac{3\sigma^2}{4\pi^2r_\ast^2}\;.
\label{eq315}
\end{align}
The approximate solutions (\ref{eq302}) and (\ref{eq306}) for $\alpha=1$ are tested against this exact solution in Figs.~\ref{fig1} and \ref{fig3}.

\begin{figure}[!t]
\includegraphics[width=0.45\textwidth]%
 {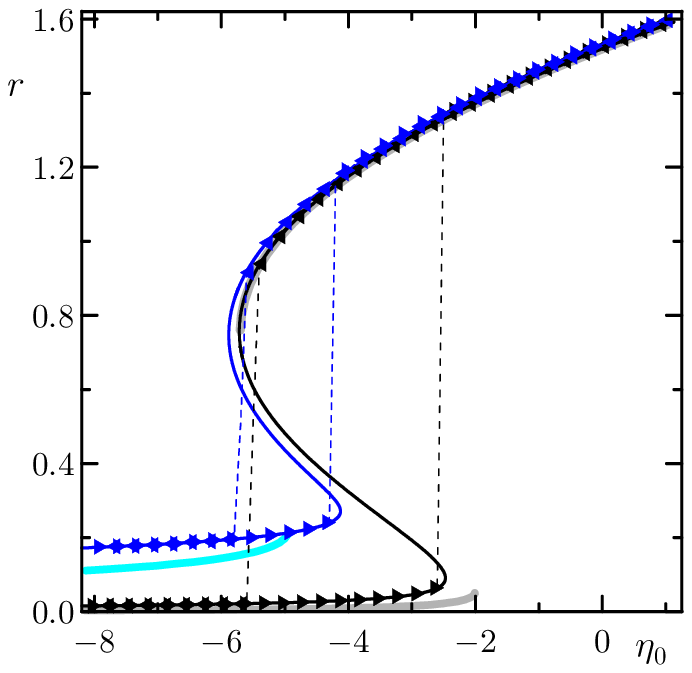}
\caption{
The results of the direct numerical simulation of the microscopic dynamics of a homogeneous population of $1000$ QIFs~(\ref{eq110},\ref{eq111}) subject to $\alpha$-stable noise are plotted with triangles for $\alpha=1/2$ (blue) and $\alpha=3/2$ (black) and the same parameter values as in Fig.~\ref{fig1}. One observes a hysteretic transition; the rightwards pointed triangles: quasiadiabatic increase of $\eta_0$, the leftward pointed triangles: quasiadiabatic decrease of $\eta_0$. The microscopic simulation results are guided by the curves of the high-precision time-independent solutions~(\ref{eqap404}) for a thermodynamic limit of $N\to\infty$; these solutions are calculated with power series expansions as described in Appendices~\ref{app:HPC05} and \ref{app:HETER}. The pale color bold lines show the approximate analytical solutions~(\ref{eq302}) and (\ref{eq307}).
}
  \label{fig2}
\end{figure}

\subsection{The exactly solvable case of a Gaussian noise}\label{ssec:QIFG}
The problem~(\ref{eq113})--(\ref{eq115}) can be also solved analytically~\cite{Volo-etal-2022} for the case of a homogeneous population $\eta=\eta_0$ with Gaussian noise, $\alpha=2$, and $I(t)=0$. Fokker--Planck equation (\ref{eq113}) with $\alpha=2$ and given $I_0$ yields time-independent firing rate (\ref{eq115})~\cite{Volo-etal-2022}
\begin{align}
r&=\sigma^{2/3}\mathcal{R}_2(A)
\nonumber\\
&=\left\{\begin{array}{cc}
\displaystyle
\frac{-9\sigma^2/(4\pi^2I_0)}{\mathrm{I}_\frac13^2(\chi_-) +\mathrm{I}_{-\frac13}^2(\chi_-) +\mathrm{I}_\frac13(\chi_-)\mathrm{I}_{-\frac13}(\chi_-)}\,,&
\displaystyle
I_0<0\,,
\\
\displaystyle
\frac{9\sigma^2/(4\pi^2I_0)}{\mathrm{J}_\frac13^2(\chi_+) +\mathrm{J}_{-\frac13}^2(\chi_+) -\mathrm{J}_\frac13(\chi_+)\mathrm{J}_{-\frac13}(\chi_+)}\,,&
\displaystyle
I_0>0\,,
\end{array}\right.
\label{eqg01}
\end{align}
where $A=I_0/\sigma^{4/3}$, $\chi_\pm=2(\pm A)^{3/2}/3$, $\mathrm{J}_n$ and $\mathrm{I}_n$ are the $n$th order Bessel function of the first kind and the modified one, respectively. The two branches of this solution (for positive and negative $I_0$) are analytic continuation of each other, but the form~(\ref{eqg01}) allows one to keep all calculations explicitly real-valued.

Substituting $I_0=\eta_0+Jr$, one finds a self-consistency problem for the firing rate $r$:
\begin{equation}
r=\sigma^{2/3}\mathcal{R}_2\left(\frac{\eta_0+Jr}{\sigma^{4/3}}\right)\,.
\label{eqg02}
\end{equation}
This equation yields dependence $r=r(\eta_0,J,\sigma)$ in a parametric form, parameterized by $A$, $-\infty<A<\infty$:
\[
r=\sigma^{2/3}\mathcal{R}_2(A)\,,
\qquad
\eta_0=\sigma^{4/3}A-J\sigma^{2/3}\mathcal{R}_2(A)\,.
\]
Similarly to the case of $\alpha\ne2$, this dependence has a multistability domain and the boundaries of this domain can be calculated from $\mathrm{d}\eta_0/\mathrm{d}r=0$:
\[
\mathrm{d}r=\sigma^{2/3}\mathcal{R}_2^\prime(A)\,\mathrm{d}A\,,
\qquad
0=\sigma^{4/3}\mathrm{d}A-J\mathrm{d}r\,,
\]
which yields
\begin{align}
J_\ast&=\frac{\sigma^{2/3}}{\mathcal{R}_2^\prime(A)}\,,
\label{eqg03}
\\
\eta_{0\ast}&=\sigma^{4/3}\left(A-\frac{\mathcal{R}_2(A)}{\mathcal{R}_2^\prime(A)}\right)\,.
\label{eqg04}
\end{align}

\subsection{Stability of macroscopic time-independent states}\label{ssec:QIFstab}
For a Cauchy noise, the stability can be analyzed rigorously in the entire parameter space. We note that Eq.~(\ref{eq118}) for $\alpha=1$ is identical to this equation without noise but with a redefined heterogeneity parameter $\Delta+\sigma\to\Delta_\sigma$. The Ott--Antonsen manifold is known to be attracting in this case.~\cite{Ott-Antonsen-2009,Mirollo-2012,Pietras-Daffertshofer-2016} Hence, for $\alpha=1$, the instabilities of macroscopic regimes can be treated within the framework of the Ott--Antonsen ansatz. Eq.~(\ref{eq123}) for $\alpha=1$ and $W_m=(\pi r-iv)\delta_{1m}$ yields
\begin{align}
\dot{r}&=\frac{\Delta_\sigma}{\pi}+2rv\;,
\label{eq316}
\\
\dot{v}&=\eta_0+Jr-\pi^2r^2+v^2\;
\label{eq317}
\end{align}
(cf Ref.~\citen{Montbrio-Pazo-Roxin-2015}).

The time-independent states $(r_0,v_0)$ are given by the equation system
\begin{align}
0&=\frac{\Delta_\sigma}{\pi}+2r_0v_0\;,
\label{eq318}
\\
0&=\eta_0+Jr_0-\pi^2r_0^2+v_0^2\;.
\label{eq319}
\end{align}
For linear perturbations of $(r_0,v_0)$, one finds the exponential growth rates
\[
\lambda=2v_0\pm\sqrt{2r_0(J-2\pi^2r_0)}\;.
\]
According to Eq.~(\ref{eq318}), $v_0=-\Delta_\sigma/(2\pi r_0)$ is always negative ($r_0$ cannot be negative by its physical meaning). If the perturbations are oscillatory (the square root in $\lambda$ is imaginary) then $\mathrm{Re}\lambda=2v_0<0$. Hence, the instability can be only monotonous. For a monotonous instability the only possible codimension 1 bifurcation is a saddle--node one, which can be always detected from the dependence of the solution on parameters, as a folding point. In Fig.~1, one can see two saddle--node bifurcation points $\mathbf{T}_\mathrm{mf}$ and $\mathbf{T}_\mathrm{nd}$ on the bold dotted curve; along the curve, the stability changes only in these points. The analysis of $\lambda$ confirms that, for all parameter values, the upper and lower branches are the branches of linearly stable regimes, while the branch near the dash-dotted line is unstable.

For $\alpha\ne1$, in the examined parameter domain, no oscillatory instabilities of time-independent macroscopic regimes were observed with direct numerical simulations. Similarly to the case of a Cauchy noise, the change of stability can be related only to saddle--node bifurcations and the upper and lower branches in Fig.~\ref{fig1} are stable.



\begin{figure}[!t]
\includegraphics[width=0.45\textwidth]%
 {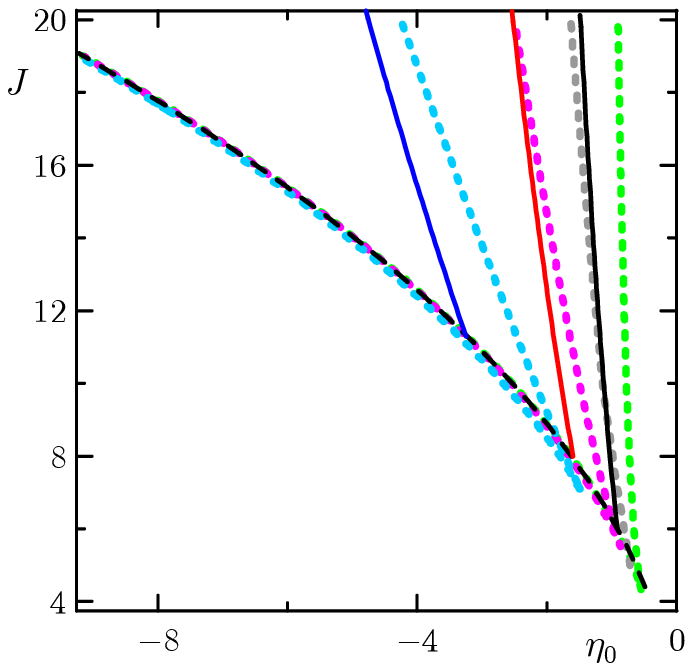}
\caption{The phase diagram of the collective regimes of a homogeneous population of QIFs~(\ref{eq110},\ref{eq111}) subject to $\alpha$-stable noise. The bifurcation curves bound the tongue-shaped bistability domain. The dotted magenta line shows the exact results (\ref{eq314},\ref{eq315}) for a Cauchy noise ($\alpha=1$): the left branch is the line of a saddle--node bifurcation of the mean-field driven regime and the right branch is the one of the noise-driven regime (see Fig.~\ref{fig1}). The dotted cyan and gray lines: numerical simulation of Eq.~(\ref{eq117}) for $\alpha=1/2$ and $3/2$, respectively. The green dotted line show the exact results (\ref{eqg03},\ref{eqg04}) for a Gaussian noise ($\alpha=2$, shown for comparison). Black dashed line: the approximate bifurcation curve~(\ref{eq304}) of the mean-field driven regime, which is identical (up to the approximation accuracy) for all $\alpha$; solid lines: the approximate bifurcation curves~(\ref{eq308},\ref{eq309}) for $\alpha=1/2$, $1$, $3/2$ (from left to right). Noise strength $\sigma=0.5$.
}
  \label{fig3}
\end{figure}

\begin{figure}[!t]
\includegraphics[width=0.45\textwidth]%
 {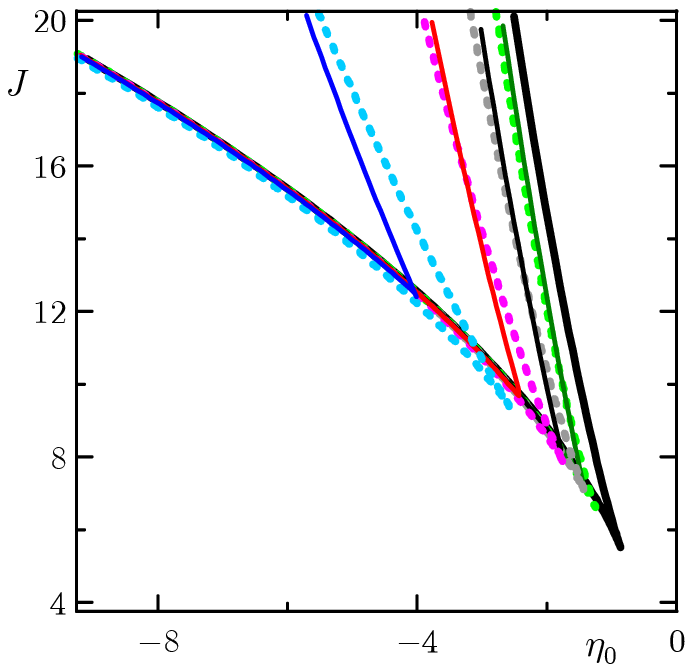}
\caption{The phase diagram of the collective regimes of a heterogeneous population of QIFs~(\ref{eq110},\ref{eq111}) subject to $\alpha$-stable noise, $\Delta=0.5$ and $\sigma=0.5$. The bifurcation curves bound the tongue-shaped bistability domain; the color coding is the same as in Fig.~\ref{fig3}: $\alpha=0.5$ (cyan/blue), $1$ (magenta/red), $1.5$ (grey/black), and $2$ (green/dark-green).
Dotted lines: exact results for Eq.~(\ref{eq117}).
Thin solid lines: theoretical bifurcation curves for approximate solution~(\ref{eq215}).
Bold black line: the domain boundary for the noise-free population.
}
  \label{fig4}
\end{figure}

\section{Discussion}
\label{sec:disc}
In the phase diagram in Fig.~\ref{fig3}, one can see that the bifurcation curve of the mean-field regime is weakly affected by the noise and its statistical properties, while the bifurcation curve of the noise-driven regime is significantly influenced by $\alpha$. The bistability domain shrinks as $\alpha$ decreases. The self-consistent perturbation theory for $\Phi(k)$ (linear in $\Phi_1$) yields reasonable accuracy even for as strong noise as $\sigma=0.5$ (see Figs.~\ref{fig1} and \ref{fig3}). For $\alpha=3/2$, the direct numerical simulation of Eq.~(\ref{eq116}) for the evolution of the characteristic function $F(k,t)$ was performed using the modification of the exponential time differencing method~\cite{Cox-Matthews-2002} for equation systems with a non-diagonal linear part.~\cite{Permyakova-Goldobin-2020} The numerical method was tested for the case $\alpha=1$ (Cauchy noise), where an analytical solution is known. For $\alpha=1/2$, a high-precision computation procedure for time-independent states was employed (see Appendix~\ref{app:HPC05}).

For the case of a heterogeneous population with a Cauchy distribution of $\eta$, the theoretical analytical approximation~(\ref{eq212}) [equivalently, Eqs.~(\ref{eq215}) and (\ref{eq217})] is examined by comparison to the `exact' numerical results in Fig.~\ref{fig4}. The analytical theory exhibits a decent accuracy even for the noise-driven regimes (the right bifurcation curve) and as large noise strength as $\sigma=0.5$. The numerical results in Fig.~\ref{fig4} are calculated with power-series expansions of time-independent $F(k)$ (Appendix~\ref{app:HETER}) with controlled accuracy $10^{-15}$. A fine detail of the mean-field driven regime can be noticed for both homo- and heterogeneous populations in Figs.~\ref{fig3} and \ref{fig4}: the results indicate that both noises with $\alpha<1$ and $\alpha>1$ slightly extend the domain of the existence of this regime towards lower excitability $\eta_0$. Both the analytical and exact left curves with $\alpha\ne1$ are shifted leftwards as compared to the case of $\alpha=1$.

The results of our analytical consideration shed light on two important issues, which we discuss below in this section.

\paragraph{Three-pseudocumulant reduction for the case of a Gaussian noise.}
For a Gaussian noise ($\alpha=2$), one can employ a pseudocumulant approach that yields an infinite chain of dynamical equations for pseudocumulants $W_m$ (\ref{eq122}). Generally, pseudocumulants form a rapidly decaying hierarchy of smallness~\cite{Goldobin-Volo-Torcini-2021}, $|W_{m+1}|/|W_m|\propto\epsilon\ll1$. Hence, one can consider finite truncations of this chain and work with low dimensional neural mass models. Obviously, in order to study the effect of noise, one must keep at least $W_2$, as $W_1$ corresponds to a wrapped Cauchy distribution of states. But is a two-pseudocumulant truncation an optimal approximation?

For a population of QIFs~(\ref{eq111},\ref{eq112}) with global synaptic coupling and Gaussian noise, the two-element reduction was reported to yield correct predictions of the thresholds of the hysteretic transitions between the time-independent regime of asynchronous firing and the regime of collective oscillations (see Fig.~1(g) in Ref.~\citen{Goldobin-Volo-Torcini-2021}). The three-element reduction was reported to reproduce all the quantitative results of the network simulation perfectly. However, there was no clear reason to include the third pseudocumulant $W_3$, but neglect higher elements.
Now, with Eq.~(\ref{eq209}), one can see that the fractional power terms start from $k^{\alpha+1}$; in Eq.~(\ref{eq118}), this term of $\Phi(k)$ emerges as an immediate response to the noise term $\sigma^\alpha|k|^\alpha$.
Meanwhile, the $k^2$-term emerges from the $C_0$-part of (\ref{eq205}); the value of constant $C_0$ was derived from the boundary condition on $\Phi(k)$ at $k\to\infty$ (\ref{eq121}).
To summarize, noise immediately enforces the $k^{\alpha+1}$-term, while the $k^2$-term is merely a result of a finer tune-up.
For $\alpha=2$, the reproduction of this scheme requires $W_3$ to be included in the analysis. Thus, a three-element reduction is the minimal model which can potentially reproduce the generic mechanism of the effect of Gaussian noise. A two-element reduction serves here as an approximation, which might yield a decent accuracy, but not an asymptotically rigorous result. This explains a dramatic enhancement of the accuracy of the three-element model compared to the two-element one in Fig.~1(g) of Ref.~\citen{Goldobin-Volo-Torcini-2021}.
Therefore, we conclude that a three-pseudocumulant reduction is a preferable default neural mass model for the case of a Gaussian noise.

\paragraph{Generalization of pseudocumulant approach for noninteger $\alpha$-stable noises.}
The fractional Fokker--Planck equation~(\ref{eq113}) or its projection into Fourier-space~(\ref{eq116}) describe the noninteger $\alpha$ case.
However, it seems to be impossible to generalize the pseudocumulant framework to solve this equation, at least, via series
$
\Phi(k>0,t)=\sum_{mn}W_{m,n}(t)k^{m+n\alpha}/(m+n\alpha)
$
and the derivation of an analog of dynamical equation system~(\ref{eq123}) for $W_{m,n}$.
First, with Eq.~(\ref{eq205}), we witnessed that the $k^{\alpha+m}$-terms are linked to the $k^{m}$-terms (with coefficient $C_0$) by the boundary condition~(\ref{eq121}) at infinity, where these two groups of terms must compensate the growing parts of each other. Generally, the latter task cannot be accomplished with a finite truncations for irrational $\alpha$.
Second, for such series we have the cases of rational $\alpha$, where terms $W_{m,n}(t)k^{m+n\alpha}$ with the same $(m+n\alpha)$ must merge. Each case of rational $\alpha=L/N$, requires individual treatment. In Appendix~\ref{app:HIA} we consider two primary cases of rational $\alpha$, $\alpha=1/2$ and $3/2$, and demonstrate that, at least in the limit of a weak noise, finite truncations of $W_{m,n}$ series fail to reproduce the impact of the $\alpha$-stable noise accurately.


\section{Conclusion}
\label{sec:conc}
We have addressed the problem of mathematical description of the macroscopic dynamics of populations of quadratic integrate-and-fire neurons subject to $\alpha$-stable white noises. The interest to these models are multifold: QIF is not only the normal form for the neuron models with the Class~I excitability near the threshold between the excitable regime and the regime of periodic spiking, but also is mathematically equivalent to the problem of the Anderson localization in one-dimensional setup~\cite{Lifshitz-Gredeskul-Pastur-1988} in condensed matter. The mathematical challenges emerging in our study are related to the specificity of QIF but not to the nature and topology of the connection network. While we report on the case of the recurrent network of chemical synaptic all-to-all connections, the formalism can be readily extended to the cases of balanced networks with sparse synaptic connections,~\cite{Volo-Torcini-2018,Volo-etal-2022} electrical synapses,~\cite{Laing-2015,Pietras-Devalle-Roxin-etal-2019,Montbrio-Pazo-2020} etc.

The research interest to non-Gaussian ($\alpha$-stable) noises in oscillator populations is additionally reinforced by the discovery that the interplay of the mechanisms of synchronization by common noise and by coupling results in non-Gaussian phase deviations $\theta_j$ from the synchronous cluster.~\cite{Goldobin-Dolmatova-2019a} In the limit of a nearly perfect synchrony (small heterogeneity of oscillators and weak individual component of noises), these deviations $\theta_j$ possess power-law tails, $\propto1/|\theta|^{2+2m}$, where $m$ is linearly proportional to the ratio of the coupling coefficient to the common noise intensity. For a repulsive coupling, $m<0$, the common noise prevails the desynchronizing action by the coupling for $m>-1/2$; the corresponding values of $\alpha=1+2m$ are in range $(0;1)$. For a non-strong attractive coupling, $0<m<1/2$, we still observe heavy power-law tails and $1<\alpha<2$. For a stronger attractive coupling the power-law tails decay non-slower than $1/|\theta|^3$, and, by virtue of the Central limiting theorem, these fluctuations generate an effective noise which converges to the Gaussian statistics. For finite populations of noise-free oscillators, the finite-size fluctuations were revealed to act as an effective common noise on the nearly identical oscillators.~\cite{Pikovsky-Dolmatova-Goldobin-2019,Peter-Gong-Pikovsky-2019} Therefore, there are generic mechanisms inducing the $\alpha$-stable component in oscillator fluctuations not only for noisy oscillator populations but also in finite ensembles of deterministic oscillators or in large ensembles with a sparse network of connections and a finite number of inbound connections for each oscillator. The macroscopic dynamics of the latter network was demonstrated to be well reproducible with an effective mean field (global) coupling endowed with an effective noise.~\cite{Volo-Torcini-2018,Goldobin-Volo-Torcini-2021,Volo-etal-2022} To summarize, one can expect abundance of oscillator (in particular, QIF) network setups where an effective noise emerge with significant $\alpha$-stable component.

The analysis of the case of fractional $\alpha$ yielded new results for the case of integer $\alpha=2$. Considering $\alpha=2$ as a limit of fractional $\alpha$ with a vanishing deviation from $2$, one sees that the minimal consistent model reduction requires three pseudocumulants $\{W_1,W_2,W_3\}$ for the Gaussian noise case, and the formal first-order correction of $W_2$ to the zeroth-order no-noise solution $W_1$ is not an asymptotically rigorous result. This sheds light on the puzzle why, for some systems, the three-pseudocumulant model reduction can provide perfect results (see the diagram in Fig.~1(g) of Ref.~\citen{Goldobin-Volo-Torcini-2021}) where the two-pseudocumulant reduction provides a reasonable but not very accurate picture. Normally, with a small parameter, the second-order correction provides an enhancement to the first-order correction, but not such a dramatic increase of accuracy. Thus, one can make a general recommendation for constructing low dimensional model reductions on the basis of pseudocumulants:~\cite{Goldobin-Volo-Torcini-2021} the rigorous asymptotic leading-order correction is given by the three-pseudocumulant model and the simplified two-pseudocumulant model can be sometimes insufficient even with a small parameter violating the applicability of the Montbri\'o--Paz\'o-Roxin Ansatz.~\cite{Montbrio-Pazo-Roxin-2015}

It turned out to be seemingly impossible to generalize the the pseudocumulant formalism to rational fractional $\alpha=L/N$ via expansions of the logarithm $\Phi(k,t)$ of the characteristic function in series with respect to $k^{1/N}$ or some other rational fractional powers of $k$. Noticeably, the circular cumulant approach~\cite{Tyulkina-etal-2018,Goldobin-Dolmatova-2019b} was reported to be useful for dealing with fractional $\alpha$-stable noises in term of oscillation phases~\cite{Dolmatova-Tyulkina-Goldobin-2023} at least in the case of an additive-in-phase noise, which is of course not our case, where the additive-in-voltage noise corresponds to a multiplicative noise in terms of the oscillation phase.

Nonetheless, for $\alpha$-stable noises, one can construct expansions of $\Phi(k,t)$ in series of the noise intensity $\sigma^\alpha$. The theoretical results derived with the latter expansion for a population of QIFs with excitatory synaptic coupling subject to non-Gaussian noise are in good agreement with the results of numerical simulation for both homogeneous (Fig.~\ref{fig3}) and heterogeneous populations (Fig.~\ref{fig4}). One observes a reasonable accuracy even for the bifurcation curves of the noise-driven regimes for as large noise amplitude as $\sigma=0.5$ (the right-hand-side branches of the cusps in Figs.~\ref{fig3} and \ref{fig4}).

\section*{ACKNOWLEDGMENTS}
The authors acknowledge the financial support from RSF (Grant no.\ 23-12-00180).

\section*{AUTHOR DECLARATIONS}
\subsection*{Conflict of Interest}
The authors have no conflicts to disclose.

\subsection*{Author Contributions}
\textbf{Denis S.\ Goldobin:}
Conceptualization (lead);
Formal analysis (equal);
Investigation (equal);
Software (equal);
Writing -- original draft (equal).
\textbf{Evelina V.\ Permyakova:}
Formal analysis (supporting);
Investigation (equal);
Software (equal);
Writing -- original draft (supporting).
\textbf{Lyudmila S.\ Klimenko:}
Conceptualization (supporting);
Formal analysis (equal);
Investigation (equal);
Writing -- original draft (equal).

\section*{DATA AVAILABILITY}
The data that support the findings of this study are available within the article and shown in the figures.

\appendix
\section{Fractional Fokker-Planck equation for additive noise}\label{app:FFP}
Here we recall the derivation of the evolution equation for the probability density function $w(\mathbf{x},t)$ for the stochastic system~(\ref{eq002}). For additive noise, one can evaluate the increment
 $\Delta w(\mathbf{x},t)=w(\mathbf{x},t+\Delta t)-w(\mathbf{x},t)$
for infinitesimal $\Delta t$;
\begin{align}
w(\mathbf{x},t+\Delta t)&=-\sum_{l=1}^{M}\frac{\partial}{\partial x_l}\Big(f_l(\mathbf{x})\,w(\mathbf{x},t)\Big)\Delta{t}
\nonumber\\
&{}+\int\limits_{-\infty}^{+\infty}\mathrm{d}\xi_{\Delta t}P(\xi_{\Delta t})\,w(\mathbf{x}-\mathbf{g}\xi_{\Delta t}\Delta t,t)\;.
\label{eq004}
\end{align}
With the characteristic function
\[
F_\mathbf{x}(\mathbf{k},t)=\langle{e^{i\mathbf{k}\cdot\mathbf{x}}}\rangle
=\int w(\mathbf{x},t)e^{i\mathbf{k}\cdot\mathbf{x}}\mathrm{d}^M\mathbf{x}\;,
\]
one can write $w(\mathbf{x},t)=(2\pi)^{-M}\int F_\mathbf{x}(\mathbf{k},t)\,e^{-i\mathbf{k}\cdot\mathbf{x}}\mathrm{d}^M\mathbf{k}$\,,
and Eq.~(\ref{eq004}) can be rewritten as
\begin{align}
&w(\mathbf{x},t+\Delta t)=-\sum_{l=1}^{M}\frac{\partial}{\partial x_l}\Big(f_l(\mathbf{x})\,w(\mathbf{x},t)\Big)\Delta{t}
\nonumber\\
&\quad
{}+\int\frac{\mathrm{d}^M\mathbf{k}}{(2\pi)^M}
\int\limits_{-\infty}^{+\infty}\mathrm{d}\xi_{\Delta t}P(\xi_{\Delta t})
F_\mathbf{x}(\mathbf{k},t)\,e^{-i\mathbf{k}\cdot\left(\mathbf{x}-\mathbf{g}\xi_{\Delta t}\Delta t\right)}
\nonumber\\
&=w(\mathbf{x},t)-\sum_{l=1}^{M}\frac{\partial}{\partial x_l}\Big(f_l(\mathbf{x})\,w(\mathbf{x},t)\Big)\Delta{t}
\nonumber\\
&\quad
{}+\frac{1}{(2\pi)^M}\int\mathrm{d}^M\mathbf{k}
\left[F_{\xi_{\Delta t}}(\mathbf{k}\cdot\mathbf{g}\Delta t)-1\right]
F_\mathbf{x}(\mathbf{k},t)\,e^{-i\mathbf{k}\cdot\mathbf{x}}\,.
\nonumber
\end{align}
For $\Delta t\to 0$, $F_{\xi_{\Delta t}}(\mathbf{k}\cdot\mathbf{g}\Delta t)=e^{\Phi_{\xi_{\Delta t}}(\mathbf{k}\cdot\mathbf{g}\Delta t)}=1+\Phi_{\xi_{\Delta t}}(\mathbf{k}\cdot\mathbf{g}\Delta t)+\mathcal{O}\left([\Phi_{\xi_{\Delta t}}(\mathbf{k}\cdot\mathbf{g}\Delta t)]^2\right)$.
Hence,
\begin{align}
&\frac{\Delta w(\mathbf{x},t)}{\Delta t}+\sum_{l=1}^{M}\frac{\partial}{\partial x_l}\Big(f_l(\mathbf{x})\,w(\mathbf{x},t)\Big)
\nonumber\\
&\qquad
{}=\int\frac{\mathrm{d}^M\mathbf{k}}{(2\pi)^M}
\frac{\Phi_{\xi_{\Delta t}}(\mathbf{k}\cdot\mathbf{g}\Delta t)}{\Delta t}
F_\mathbf{x}(\mathbf{k},t)\,e^{-i\mathbf{k}\cdot\mathbf{x}}\,.
\label{eq005}
\end{align}
With $\xi_{\Delta t}\propto(\Delta t)^{1/\alpha-1}$, and noise of amplitude $\sigma$, one should take $c=(\Delta t)^{1/\alpha-1}\sigma$ and find
\begin{equation}
\dot\Phi_t^{(\xi)}(\mathbf{k}\cdot\mathbf{g})\equiv
\frac{\Phi_{\xi_{\Delta t}}(\mathbf{k}\cdot\mathbf{g}\Delta t)}{\Delta t}=-\sigma^\alpha|\mathbf{g}\cdot\mathbf{k}|^\alpha
\label{eq006}
\end{equation}
(here $\mu=\beta=0$).
Eq.~(\ref{eq005}) with~(\ref{eq006}) is identical to~(\ref{eq105}).

\section{Probability distribution for a nonsmooth characteristic function}
\label{app:PDF}
With (\ref{eq210}), one can find the following characteristic function with derivative discontinuity at $k=0$:
\begin{align}
&F(k)=e^{\Phi}=e^{-W_{1,0\mathrm{r}}|k|-iW_{1,0\mathrm{i}}k}
\nonumber\\
&\qquad
\times e^{-W_{2,0\mathrm{r}}\frac{k^2}{2}-iW_{2,0\mathrm{i}}\frac{k|k|}{2} -W_{1,1\mathrm{r}}\frac{|k|^{\alpha+1}}{\alpha+1}-iW_{1,1\mathrm{i}}\frac{|k|^\alpha k}{\alpha+1}-\dots}
\nonumber\\
&
=\left[1-W_{2,0\mathrm{r}}\frac{k^2}{2}-iW_{2,0\mathrm{i}}\frac{k|k|}{2} -W_{1,1\mathrm{r}}\frac{|k|^{\alpha+1}}{\alpha+1}
\right.
\nonumber\\
&\qquad\left.
-iW_{1,1\mathrm{i}}\frac{|k|^\alpha k}{\alpha+1}-\dots\right]e^{-W_{1,0\mathrm{r}}|k|-iW_{1,0\mathrm{i}}k}\;.
\nonumber
\end{align}
The inverse Fourier transform~(\ref{eq107}) yields
\begin{align}
&w(V)=\int\limits_{-\infty}^{+\infty}
 \frac{\mathrm{d}k}{2\pi}
\bigg(1-W_{2,0\mathrm{r}}\frac{k^2}{2}-iW_{2,0\mathrm{i}}\frac{k|k|}{2}
\nonumber\\
&{}
 -W_{1,1\mathrm{r}}\frac{|k|^{\alpha+1}}{\alpha+1}
 -iW_{1,1\mathrm{i}}\frac{|k|^\alpha k}{\alpha+1}
 -\dots\bigg)
 e^{-W_{1,0\mathrm{r}}|k|-i(W_{1,0\mathrm{i}}+V)k}
\nonumber\\
&=\int\limits_{-\infty}^{+\infty}
 \frac{\mathrm{d}k}{2\pi}
\bigg(1+\frac{W_{2,0\mathrm{r}}}{2}\frac{\partial^2}{\partial V^2}
 +\frac{W_{2,0\mathrm{i}}}{2}\frac{\partial}{\partial V}\left|\frac{\partial}{\partial W_{1,0\mathrm{r}}}\right|
\nonumber\\
&\quad{}
 +\frac{W_{1,1\mathrm{r}}}{\alpha+1}\frac{\partial^2}{\partial V^2}\left|\frac{\partial}{\partial W_{1,0\mathrm{r}}}\right|^{\alpha-1}
 +\frac{W_{1,1\mathrm{i}}}{\alpha+1}\frac{\partial}{\partial V}\left|\frac{\partial}{\partial W_{1,0\mathrm{r}}}\right|^{\alpha}
\nonumber\\
&\qquad\qquad
 +\dots\bigg)
 e^{-W_{1,0\mathrm{r}}|k|-i(W_{1,0\mathrm{i}}+V)k}
\nonumber\\
&=
\bigg(1+\frac{W_{2,0\mathrm{r}}}{2}\frac{\partial^2}{\partial V^2}
 +\frac{W_{2,0\mathrm{i}}}{2}\frac{\partial}{\partial V}\left|\frac{\partial}{\partial W_{1,0\mathrm{r}}}\right|
\nonumber\\
&\qquad{}
 +\frac{W_{1,1\mathrm{r}}}{\alpha+1}\frac{\partial^2}{\partial V^2}\left|\frac{\partial}{\partial W_{1,0\mathrm{r}}}\right|^{\alpha-1}
 +\frac{W_{1,1\mathrm{i}}}{\alpha+1}\frac{\partial}{\partial V}\left|\frac{\partial}{\partial W_{1,0\mathrm{r}}}\right|^{\alpha}
\nonumber\\
&\qquad\qquad
 +\dots\bigg)
 \frac{\pi^{-1}W_{1,0\mathrm{r}}}{W_{1,0\mathrm{r}}^2+(V+W_{1,0\mathrm{i}})^2}\;.
\label{eqappB1}
\end{align}

\paragraph{Firing rate.}
With distribution~(\ref{eqappB1}), one can calculate firing rate~(\ref{eq115}). For large $V$, the differentiation $|\partial/\partial W_{1,0\mathrm{r}}|$ does not decrease the exponent of the power law of a tail $\propto1/V^2$ and the differentiation $\partial/\partial V$ turns $\propto 1/V^n$ into $\propto 1/V^{n+1}$; therefore,
\begin{equation}
r=\lim\limits_{V\to\pm\infty}V^2w(V)=\frac{\mathrm{Re}(W_{1,0})}{\pi}\;,
\label{eqappB2}
\end{equation}
and this result holds with any number of corrections $W_{m,n}$.

\paragraph{Mean value $\langle{V}\rangle$.}
The mean value of $w(V)$ is $-W_{1,0\mathrm{i}}$ and also not affected by the higher order corrections $W_{m,n}$. Indeed, using partial integration, one can calculate the integral
\begin{align}
&\langle{V}\rangle
=\mathrm{P.V.}\int\limits_{-\infty}^{+\infty} \mathrm{d}V
\bigg(V+V\frac{W_{2,0\mathrm{r}}}{2}\frac{\partial^2}{\partial V^2}
\nonumber\\
&\quad{}
 +V\frac{W_{2,0\mathrm{i}}}{2}\frac{\partial}{\partial V}\left|\frac{\partial}{\partial W_{1,0\mathrm{r}}}\right|
 +V\frac{W_{1,1\mathrm{r}}}{\alpha+1}\frac{\partial^2}{\partial V^2}\left|\frac{\partial}{\partial W_{1,0\mathrm{r}}}\right|^{\alpha-1}
\nonumber\\
&\qquad{}
 +V\frac{W_{1,1\mathrm{i}}}{\alpha+1}\frac{\partial}{\partial V}\left|\frac{\partial}{\partial W_{1,0\mathrm{r}}}\right|^{\alpha}
 +\dots\bigg)
 w_\mathrm{LD}(V)
\nonumber\\
&\quad{}
=\mathrm{P.V.}\int\nolimits_{-\infty}^{+\infty}V w_\mathrm{LD}(V)\,\mathrm{d}V
 -\frac{W_{2,0\mathrm{r}}}{2}w_\mathrm{LD}(V)\Big|_{-\infty}^{+\infty}
\nonumber\\
&\qquad\quad{}
-\frac{W_{2,0\mathrm{i}}}{2}\left|\frac{\partial}{\partial W_{1,0\mathrm{r}}}\right|\int\nolimits_{-\infty}^{+\infty} w_\mathrm{LD}(V)\,\mathrm{d}V
\nonumber\\
&\qquad\quad{}
 -\frac{W_{1,1\mathrm{r}}}{\alpha+1}\left|\frac{\partial}{\partial W_{1,0\mathrm{r}}}\right|^{\alpha-1}w_\mathrm{LD}(V)\Big|_{-\infty}^{+\infty}
\nonumber\\
&\qquad\quad{}
 -\frac{W_{1,1\mathrm{i}}}{\alpha+1}\left|\frac{\partial}{\partial W_{1,0\mathrm{r}}}\right|^{\alpha}\int\nolimits_{-\infty}^{+\infty} w_\mathrm{LD}(V)\,\mathrm{d}V
 +\dots
 \nonumber\\
&\quad{}
 =-\mathrm{Im}(W_{1,0})\;,
\label{eqappB3}
\end{align}
where we used the conditions $\int_{-\infty}^{+\infty} w_\mathrm{LD}(V)\,\mathrm{d}V=1$ and $w_\mathrm{LD}(\pm\infty)=0$.

\section{Expansions for half-integer $\alpha$}\label{app:HIA}
Let us first make a general observation about $\Phi(k)$ and Eq.~(\ref{eq118}). Function $\Phi(k)$ vanishes at $k=0$ and thus its series can contain only positive powers of $k$.
In Eq.~(\ref{eq118}), the term $k(\partial^2\Phi/\partial k^2)$ decreases the degree of any term by one and leads to inadmissible negative degrees for noninteger power exponents if only these terms are not canceled at some order. The noise term $\sigma^\alpha|k|^\alpha$ forces terms with noninteger power exponents which have to be canceled for some lower degrees.

\subsection{Expansion for $\alpha=1/2$}
For $\alpha=1/2$, we consider
\[
\Phi(k>0)=\sum_{n=2}^{+\infty}W_\frac{n}{2}\frac{k^\frac{n}{2}}{n/2}\,,
\]
where term $W_\frac{1}{2}=0$ is inadmissible, as it induces terms with negative power exponents.
For this series Eq.~(\ref{eq118}) yields in the orders $k^{1/2}$, $k^1$, $k^{3/2}$, $k^2$, $k^{5/2}$, and $k^3$:
\\
\begin{subequations}
\label{eqap12}
\begin{align}
0&=i\frac12 W_\frac32-\sigma^\alpha \;,
\label{eqap12a}
\\
\dot W_1&=\Delta -iI_0+i\left(-W_2+W_1^2\right)\;,
\label{eqap12b}
\\
\dot W_\frac32&=i\frac32\left(-\frac{3}{2}W_\frac52+2W_1W_\frac32\right) \;,
\label{eqap12c}
\\
\dot W_2&=2i\left(-2W_3+(W_\frac32)^2+2W_1W_2\right)\;,
\label{eqap12d}
\\
\dot W_\frac52&=i\frac52\left(-\frac{5}{2}W_\frac72+2W_1W_\frac52+2W_2W_\frac32\right)\;,
\label{eq3ap12e}
\\
\dot W_3&=3i\left(-3W_4+W_2^2+2W_1W_3+2W_\frac32 W_\frac52\right) \;,
\label{eqap12f}
\\
&\dots\;.
\nonumber
\end{align}
\end{subequations}

In the odd lines of the latter equation system we find algebraic equations but not the dynamical ones. Indeed, in (\ref{eqap12a}) we find $W_\frac32=-i2\sigma^\alpha$; in (\ref{eqap12c}) the variable $W_\frac32$ and its derivative are already superimposed by (\ref{eqap12a}) and we have expression for $W_\frac52=(4/3)W_1W_\frac32=-i8\sigma^\alpha W_1/3$; and so forth.
Therefore, for $\alpha=1/2$, one cannot write down an ODE systems and immediately run it numerically. Instead, one should deal with some of these ODEs carefully and perform direct numerical integration only for the integer-order dynamical equations~(\ref{eqap12b}), (\ref{eqap12d}), (\ref{eqap12f}), etc. The former ODEs should be solved as algebraic equations, where some terms can be time-derivatives of the integer-order elements.
For $\alpha=1/2$, the leading noise-induced term $W_\frac32$ appears for the first time in the integer-order equations in (\ref{eqap12d}) for the evolution of $W_2$.

In the limit of a weak noise $\sigma\ll1$, any truncated expansion~(\ref{eqap12}) is inconsistent with solution~(\ref{eq209}).
Indeed, according to Eqs.~(\ref{eqap12}) with zero time-derivatives $\dot{W}_\frac{n}{2}$, the noninteger elements impact the integer ones only via quadratic terms of the form $W_\frac{2n+1}{2}W_\frac{2m+1}{2}$. All noninteger elements $W_\frac{2n+1}{2}$ are proportional to $\sigma^\alpha$ and can have higher-order contributions $\propto\sigma^{3\alpha}$, etc. Hence, $W_\frac{2n+1}{2}W_\frac{2m+1}{2}\propto\sigma^{2\alpha}$ and can have contributions $\propto\sigma^{4\alpha}$, etc., which means that the integer-order elements have corrections starting from $\propto\sigma^{2\alpha}$ and the leading $\sigma^\alpha$-correction [see Eq.~(\ref{eq209})] is lost.

\subsection{Expansion for $\alpha=3/2$}
For $\alpha=3/2$, we consider
\[
\Phi(k>0)=W_1k+\sum_{n=4}^{+\infty}W_\frac{n}{2}\frac{k^\frac{n}{2}}{n/2}\,,
\]
where terms $W_\frac{1}{2}=W_\frac{3}{2}=0$ are inadmissible, as they induce terms with negative power exponents. For this series Eq.~(\ref{eq118}) yields in the orders $k^1$, $k^{3/2}$, $k^2$, \dots:
\\
\begin{subequations}
\label{eqap32}
\begin{align}
\dot W_1&=\Delta -iI_0 +i\left(-W_2+W_1^2\right)\;,
\label{eqap32a}
\\
0&=i\frac32 W_\frac52-\sigma^\alpha \;,
\label{eqap32b}
\\
\dot W_2&=4i\left(-W_3+W_1W_2\right)\;,
\label{eqap32c}
\\
\dot W_\frac52&=i\frac52\left(-\frac{5}{2}W_\frac72+2W_1W_\frac52\right)\;,
\label{eqap32d}
\\
\dot W_3&=3i\left(-3W_4+W_2^2+2W_1W_3\right) \;,
\label{eqap32e}
\\
\dot W_\frac72&=\dots\;,
\nonumber
\\
\dot W_4&=4i\left(-4W_5+(W_\frac52)^2+2W_1W_4+2W_2W_3\right) \;,
\label{eqap32f}
\\
&\dots\;.
\nonumber
\end{align}
\end{subequations}

Notice, in the even lines of the latter equation system we again find algebraic equations but not the dynamical ones. Indeed, in (\ref{eqap32b}) we find $W_\frac52=-i(2/3)\sigma^\alpha$; in (\ref{eqap32d}) the variable $W_\frac52$ and its derivative are already superimposed by (\ref{eqap32b}) and we have expression for $W_\frac72=(4/5)W_1W_\frac52=-i8\sigma^\alpha W_1/15$; and so forth.
Therefore, for rational noninteger $\alpha$, one cannot write down an ODE systems and immediately run it numerically. Instead, one should perform direct numerical integration only for the integer-order dynamical equations~(\ref{eqap32a}), (\ref{eqap32c}), (\ref{eqap32e}), (\ref{eqap32f}), etc.,
and deal with the rest of equations carefully (solve them as algebraic equations, where some terms can be time-derivatives of the integer-order elements). Remarkably, for $\alpha=3/2$, the leading noise-induced term $W_\frac52$ appears for the first time in the integer-order equations only in (\ref{eqap32f}) for the evolution of $W_4$.

In the limit of a weak noise $\sigma\ll1$, any truncated expansion~(\ref{eqap32}) is inconsistent with solution~(\ref{eq209}). As well as for $\alpha=1/2$, truncated equation systems yield the integer-order elements with corrections starting from $\propto\sigma^{2\alpha}$ and the leading $\sigma^\alpha$-correction [see Eq.~(\ref{eq209})] is lost.

\section{High-precision computation of a~time-independent state for $\alpha=1/2$}
\label{app:HPC05}
For the time-independent states of a homogeneous population ($\Delta=0$), the characteristic function evolution equation~(\ref{eq117}) simplifies to the problem
\begin{equation}
\frac{\mathrm{d}^2F}{\mathrm{d}k^2}-I_0F -i\sigma^\alpha k^{\alpha-1} F=0\;,
\quad k>0\;,
\label{eqap401}
\end{equation}
where $I_0=\eta_0+Jr$ and $r=-\pi^{-1}\mathrm{Re}\left(\frac{\mathrm{d}F}{\mathrm{d}k}|_{k=+0}\right)$. We consider only the case of a homogeneous population here, since a finite heterogeneity $\Delta$ lifts the convergence issues and a plain direct numerical simulation of~(\ref{eq117}) can be carried out with a modified exponential time differencing method~\cite{Cox-Matthews-2002,Permyakova-Goldobin-2020} for all $\alpha$.

Consider an excitable state (noise-driven regime) $I_0<0$ and $k\to\infty$ for $\alpha<1$. One finds $F\propto e^{\pm i\sqrt{-I_0}k}$ as $\sigma^\alpha k^{\alpha-1}\to 0$; the characteristic function does not decay for large $k$. This is expectable, since the probability distribution of the noise-free population with $I_0<0$ is $w(V)=\delta(V+\sqrt{-I_0})$ and
\begin{equation}
\lim\limits_{\sigma\to0}F(k)=\lim\limits_{\sigma\to0}\langle e^{ikV}\rangle=e^{-ik\sqrt{-I_0}}
\quad\mbox{ for }I_0<0\;,
\label{eqap402}
\end{equation}
and the noise term vanishes in (\ref{eqap401}) for large $k$ and $\alpha<1$. In practice, for $\sigma\ne0$ the characteristic function decays but not exponentially fast. For an exponentially fast decay of $F(k)$ with $k$ ($\alpha>1$) we simulated problem~(\ref{eq117}) within a finite domain $0\le k\le L_k$ with $L_k$ providing $|F(L_k)|<10^{-15}$.
For $\alpha<1$, the decay is exponentially fast only for $I_0>0$ and the cases of positive and negative $I_0$ should be handled separately.

First, for time-independent states, we find the scaling laws with respect to the noise strength $\sigma$. With rescaling $k\to\kappa/\sigma^\frac{\alpha}{\alpha+1}$, one can recast (\ref{eqap401}) as
\begin{equation}
\frac{\mathrm{d}^2F}{\mathrm{d}\kappa^2}-AF -i\kappa^{\alpha-1} F=0\;,
\quad \kappa>0\;,
\label{eqap403}
\end{equation}
where $A=I_0/\sigma^\frac{2\alpha}{\alpha+1}$ and
$F=1+(-\pi r+iv)\kappa/\sigma^\frac{\alpha}{\alpha+1}+\dots\equiv1+(-\pi\mathcal{R}_\alpha+i\mathcal{V}_\alpha)\kappa+\dots$\,. Hence, the firing rate
\begin{equation}
r=\sigma^\frac{\alpha}{\alpha+1} \mathcal{R}_\alpha\left(\frac{I_0}{\sigma^\frac{2\alpha}{\alpha+1}}\right)
\label{eqap404}
\end{equation}
[cf (\ref{eqg02}) for $\alpha=2$].
Second, we restrict our consideration in this section to the case $\alpha=1/2$, where arbitrary precision of computations can be achieved.

For $A>0$, the characteristic function is not only exponentially fast decaying but also its power series around $\kappa=0$
\begin{equation}
F(\kappa>0)=\sum_{n=0}^{\infty}a_n\kappa^n+\sum_{n=1}^{\infty}b_n\kappa^{n+\frac12}
\label{eqap405}
\end{equation}
converges. Here one sets $a_0=1$, complex $a_1=-\mathcal{W}_1$ and finds from (\ref{eqap403}) $b_0=0$, $b_1=4i/3$, and
\begin{equation}
a_{n+2}=\frac{Aa_n+ib_n}{(n+2)(n+1)}\;,
\quad
b_{n+2}=\frac{Ab_n+ia_{n+1}}{(n+5/2)(n+3/2)}
\label{eqap406}
\end{equation}
for $n\ge0$. Expansion~(\ref{eqap405}) with coefficients (\ref{eqap406}) acquires the form
\begin{equation}
F(\kappa>0)=F_0(\kappa)+\mathcal{W}_1F_1(\kappa)\,,
\label{eqap407}
\end{equation}
where functions $F_0(\kappa)$ and $F_1(\kappa)$ grow exponentially fast with $\kappa$ but finite complex coefficient
\begin{equation}
\mathcal{W}_1=-\lim\limits_{\kappa\to+\infty}\frac{F_0(\kappa)}{F_1(\kappa)}
\label{eqap408}
\end{equation}
provides an exponentially fast decay for the sum  $F(\kappa)$ (\ref{eqap407}). The series~(\ref{eqap405}) truncated at $N=40$ allows for computation of $\mathcal{R}_{1/2}(A)$ with error below $10^{-15}$.

Noticeably, with $A<0$ but close to zero, large finite $\kappa$ in (\ref{eqap403}) still yield an exponentially fast decay $F(\kappa)\propto\exp\big[{-i\sqrt{-A-i/\sqrt{\kappa}}\kappa}\big]$. In practice, series~(\ref{eqap405}) converges for $\kappa$ large enough to provide the accuracy $10^{-15}$ as far as for $A\gtrsim-0.5$ and $10^{-5}$ for $A\gtrsim-1$, but not only for positive $A$. Accurate computations with this procedure below $A\approx-1$ are impossible.

\begin{figure}[!t]
\includegraphics[width=0.45\textwidth]%
 {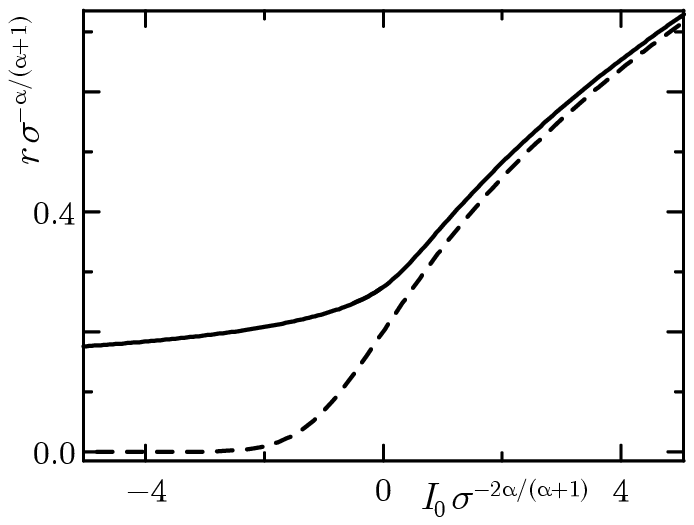}
\caption{Firing rate functions $\mathcal{R}_\alpha(A)$ for time-independent states of homogeneous populations of QIFs subject to noise are plotted with the solid line for $\alpha=1/2$ and with the dashed line for $\alpha=2$ (\ref{eqg01}).
}
  \label{fig5}
\end{figure}

For $A<0$, the convergence radius of series (\ref{eqap405}) for $F_0(\kappa)$ and $F_1(\kappa)$ becomes quite short and the series form of $F_0(\kappa)$ and $F_1(\kappa)$ becomes insufficient for accurate computation of the limit~(\ref{eqap408}). We calculate $F_0(\kappa_0)$, $F_0^\prime(\kappa_0)$ and $F_1(\kappa_0)$, $F_1^\prime(\kappa_0)$ for $\kappa_0=0.5$ and use them as the initial conditions for the numerical integration of the ordinary differential equation (\ref{eqap403}) by a 4th order Runge--Kutta method. Notice, this numerical integration of (\ref{eqap403}) from $\kappa=0$ is impossible due to the divergence of the coefficient of the last term for $\alpha<1$. The accumulation of the numeric error does not allow for such integration far beyond $L_k=60/(1+\sqrt{-A})$ (the error is kept below $10^{-15}$).

Further, we have to calculate the asymptotic behavior of $F(\kappa)$ at large $\kappa$. This can be better done in terms of the logarithm of the characteristic function $\Phi(\kappa)$; the time-independent solution of (\ref{eq118}) obeys
\[
\Phi^{\prime\prime}+(\Phi^\prime)^2+\varkappa^2-\frac{i}{\sqrt{\kappa}}=0\,,
\]
where $A\equiv-\varkappa^2$. According to Eq.~(\ref{eqap402}), the required asymptotic solution $\Phi^\prime(\kappa)\to -i\varkappa$; therefore, we consider perturbed $\Phi^\prime(\kappa>0)=-i\varkappa+\psi(\kappa)$:
\[
\psi^\prime+\psi^2-2i\varkappa\psi-\frac{i}{\sqrt{\kappa}}=0\,.
\]
For this equation one can find the asymptotic expansion of $\psi$ with the following converging iterative procedure:
\[
\psi_{n+1}=\frac{\psi_n^\prime+(\psi_n)^2}{2i\varkappa}-\frac{1}{2\varkappa\sqrt{\kappa}}\;.
\]
Starting with $\psi_0=0$, the $n$th iteration of this mapping yields the expansion of $\psi(\kappa)$ up to the term $\propto1/\kappa^{n/2}$. Approximation $\psi_{N=20}$ with terms beyond $1/\kappa^{N/2}$ dropped yields $\psi(L_k)$ with error below $10^{-15}$ uniformly over negative $A$. Finally, we can compute $\mathcal{W}_1$ from
\[
\frac{F_0^\prime(L_k)+\mathcal{W}_1F_1^\prime(L_k)}{F_0(L_k)+\mathcal{W}_1F_1(L_k)} =-i\varkappa+\psi(L_k)
\]
and $\mathcal{R}_{1/2}=\pi^{-1}\mathrm{Re}\mathcal{W}_1$.

With the firing rate~(\ref{eqap404}), similarly to the case of Gaussian noise in Sec.~\ref{ssec:QIFG}, one obtains the bifurcation curves from the condition $\mathrm{d}\eta_0/\mathrm{d}r=0$:
\begin{align}
J_\ast&=\frac{\sigma^{1/3}}{\mathcal{R}_{1/2}^\prime(A)}\,,
\label{eqap409}
\\
\eta_{0\ast}&=\sigma^{2/3}\left(A-\frac{\mathcal{R}_{1/2}(A)}{\mathcal{R}_{1/2}^\prime(A)}\right)\,.
\label{eqap410}
\end{align}
Computed dependence $\mathcal{R}_{1/2}(A)$ plotted in Fig.~\ref{fig5} yields the bifurcation curves plotted in Fig.~\ref{fig3} with the blue dashed line.

\section{Power-series expansion of characteristic function for time-independent states of heterogeneous populations and $\alpha=n/2$}\label{app:HETER}
For $\alpha=1/2$ and $\Delta\ne 0$, Eq.~(\ref{eq117}) yields a straightforward modification of expansion~(\ref{eqap405},\ref{eqap406}) for time-independent states:
\begin{align}
F(\kappa>0)&=\sum_{n=0}^{\infty}a_n\kappa^n+\sum_{n=1}^{\infty}b_n\kappa^{n+\frac12}
\label{eqap501}
\\
a_0=1\,,&\quad
a_1=-\mathcal{W}_1\,,\quad
b_0=0\,,\quad
b_1=\frac{4i}{3}\,,
\label{eqap502}
\\
n\ge0\,:&\quad
a_{n+2}=\frac{\left[A+\frac{i\Delta}{\sigma^\frac{2\alpha}{\alpha+1}}\right]a_n +ib_n}{(n+2)(n+1)}\;,
\label{eqap503}
\\
n\ge0\,:&\quad
b_{n+2}=\frac{\left[A+\frac{i\Delta}{\sigma^\frac{2\alpha}{\alpha+1}}\right]b_n +ia_{n+1}}{(n+5/2)(n+3/2)}\;.
\label{eqap504}
\end{align}
For $\alpha=3/2$, one finds
\begin{align}
F(\kappa>0)&=\sum_{n=0}^{\infty}a_n\kappa^n+\sum_{n=2}^{\infty}b_n\kappa^{n+\frac12}
\label{eqap505}
\\
a_0=1\,,&\quad
a_1=-\mathcal{W}_1\,,\quad
a_2=\frac{A}{2}+\frac{i\Delta}{2\sigma^\frac{2\alpha}{\alpha+1}}\,,
\label{eqap506}
\\
&b_0=b_1=0\,,\quad
b_2=\frac{4i}{15}\,,
\label{eqap507}
\\
n\ge1\,:&\quad
a_{n+2}=\frac{\left[A+\frac{i\Delta}{\sigma^\frac{2\alpha}{\alpha+1}}\right]a_n +ib_{n-1}}{(n+2)(n+1)}\;,
\label{eqap508}
\\
n\ge1\,:&\quad
b_{n+2}=\frac{\left[A+\frac{i\Delta}{\sigma^\frac{2\alpha}{\alpha+1}}\right]b_n +ia_{n}}{(n+5/2)(n+3/2)}\;.
\label{eqap509}
\end{align}
For $\alpha=2$ (Gaussian noise), we have an analytical solution for a homogeneous population, but the Residue theorem is unapplicable for analytic function~(\ref{eqg01}) and we do not have an exact analytical solution for $\Delta\ne0$. Eq.~(\ref{eq117}) yields for time-independent states:
\begin{align}
F(\kappa>0)&=\sum_{n=0}^{\infty}a_n\kappa^n
\label{eqap510}
\\
a_0=1\,,&\quad
a_1=-\mathcal{W}_1\,,\quad
a_2=\frac{A}{2}+\frac{i\Delta}{2\sigma^\frac{2\alpha}{\alpha+1}}\,,
\label{eqap511}
\\
n\ge1\,:&\quad
a_{n+2}=\frac{\left[A+\frac{i\Delta}{\sigma^\frac{2\alpha}{\alpha+1}}\right]a_n +ia_{n-1}}{(n+2)(n+1)}\;.
\label{eqap512}
\end{align}


\section{Direct numerical simulation of the microscopic dynamics of population of QIFs}\label{app:microDNS}
The simulation of microscopic dynamics of QIFs~(\ref{eq110},\ref{eq111}) (Fig.~\ref{fig2}) was performed with the Euler method, as no higher-order stochastic Runge--Kutta schemes are available for non-Gaussian white noises. The voltage reset threshold $B=1000$, time stepsize $\Delta t=5\cdot10^{-5}$. The iteration for one time step was conducted in two stages:

At the first (dynamical) stage, we calculated
$$
V_j(t+\Delta t):=V_j(t)+\left(\eta_j+V_j^2(t)\right)\Delta t +\sigma(\Delta t)^{1/\alpha}\zeta_j\,,
$$
where independent identically distributed symmetric ($\beta=0$) $\alpha$-stable random variables $\zeta_j$ with scale $c=1$ were generated by means of the following formula~\cite{Chambers-etal-1976,Misiorek-Weron-2012}:
\begin{equation}
\zeta_j=\left\{
\begin{array}{cc}
\tan{R_1}\,,&\mbox{ for }\alpha=1\,,\\
\frac{\sin{\alpha R_1}}{(\cos{R_1})^{1/\alpha}}
\left(\frac{\cos(R_1-\alpha R_1)}{-\ln{R_2}}\right)^\frac{1-\alpha}{\alpha}\,,&\mbox{ for }\alpha\ne1\,,
\end{array}
\right.
\end{equation}
the auxiliary random number $R_1$ is uniformly distributed in open interval $(-\pi/2;\pi/2)$ and $R_2$ is in $(0;1)$.
Importantly, the build-in pseudorandom number generators (of FORTRAN, C, etc.) have too short period, which is marginally sufficient for $\alpha=1$ or $2$ but dramatically insufficient for fractional values of $\alpha$ when the statistical convergence is slower. We employed the pseudorandom number generator MT19937~\cite{Matsumoto-Nishimura-1998} with declared period $2^{19937}$.

At the second (resetting) stage, we counted the number $N_B$ of QIFs with $V_j>B$ and reset these $V_j$ to $-B$, then shifted the states of all QIFs $V_j(t+\Delta t):=V_j(t+\Delta t)+JN_B/N$, and repeated the procedure until $N_B=0$.

For quasiadiabatic parameter variation, we simulated the population for a transient period of length $100$ and computed the average values over the next period of length $400$; the final population state was used as an initial state for the next set of parameters.

\end{document}